\documentclass[
aps,
prd,
floats,
floatfix,
twocolumn,
superscriptaddress,
nofootinbib,
noshowpacs]{revtex4-2}

\usepackage{hyperref}
\usepackage{amssymb}
\usepackage{amsmath}
\usepackage{physics}
\usepackage{latexsym}

\usepackage{verbatim}
\usepackage{mathrsfs}
\usepackage{amsfonts}

\usepackage{graphicx,subfigure}
\usepackage{epsfig}

\usepackage{units}
\usepackage{overpic}
\usepackage[utf8]{inputenc}
\usepackage{rotating}
\usepackage{enumerate}

\usepackage{soul}

\usepackage{xfrac}
\usepackage{xcolor}
\usepackage{multirow}

\begin{document}

\title{Linear stability of nonrelativistic selfinteracting boson stars}
\author{Emmanuel Ch\'avez Nambo}
\affiliation{Instituto de F\'isica y Matem\'aticas,
Universidad Michoacana de San Nicol\'as de Hidalgo,
Edificio C-3, Ciudad Universitaria, 58040 Morelia, Michoac\'an, M\'exico}
\author{Alberto Diez-Tejedor}
\affiliation{Departamento de Física, División de Ciencias e Ingenierías, Campus León, Universidad de Guanajuato, C.P. 37150, León, México}
\author{Armando A. Roque}
\affiliation{Unidad Acad\'emica de F\'isica, Universidad Aut\'onoma de Zacatecas, 98060 Zacatecas, M\'exico.}
\affiliation{Departamento de Física, División de Ciencias e Ingenierías, Campus León, Universidad de Guanajuato, C.P. 37150, León, México}
\author{Olivier Sarbach}
\affiliation{Instituto de F\'isica y Matem\'aticas,
Universidad Michoacana de San Nicol\'as de Hidalgo,
Edificio C-3, Ciudad Universitaria, 58040 Morelia, Michoac\'an, M\'exico}

\date{\today}

\begin{abstract}
In this paper we study the linear stability of selfinteracting boson stars in the nonrelativistic limit of the Einstein-Klein-Gordon theory. For this purpose, based on a combination of analytic and numerical methods, we determine the behavior of general linear perturbations around the stationary and spherically symmetric solutions of the Gross-Pitaevskii-Poisson system. In particular, we conclude that ground state configurations are linearly stable if the selfinteraction is repulsive, whereas there exist a state of maximum mass that divides the stable and the unstable branches in case the selfinteraction is attractive. 
Regarding the excited states, they are in general unstable under generic perturbations, although we identify a stability band in the first excited states of the repulsive theory. This result is independent of the mass of the scalar field and the details of the selfinteraction potential, and it is in contrast to the situation of vanishing selfinteraction, in which excited states are always unstable.
\end{abstract}

\maketitle

\section{Introduction}

The Einstein-Klein-Gordon system constitutes a nonlinear field theory that allows regular and localized configurations that do not disperse in time~\cite{Kaup:1968zz, Ruffini:1969qy}. These configurations are usually referred to as boson stars~\cite{Jetzer:1991jr, Lee:1991ax, Liddle:1992fmk, Schunck:2003kk, Liebling:2012fv, Zhang:2018slz, Visinelli:2021uve}, and their theoretical existence is possible due to the equilibrium between the repulsive ``pressure'' of the scalar field and the attractive nature of the gravitational interaction. In its simplest realization the Klein-Gordon equation consists only of the kinetic and the mass term, although boson star solutions also exist if selfinteractions are included~\cite{Colpi:1986ye, Schunck:1999zu, 2011PhRvD..84d3531C,2011PhRvD..84d3532C, Eby:2015hsq,Chavanis:2017loo,Schiappacasse:2017ham, Siemonsen:2020hcg,Chavanis:2022fvh,Jain:2023tsr}.

The phenomenological relevance of boson stars depends crucially on their stability properties.  Spherically symmetric ground state configurations are known to consist of a stable and an unstable branch~\cite{Gleiser:1988rq} divided by the state of maximum mass~\cite{Gleiser:1988ih}. This holds true in absence of selfinteractions, as well as for the theory with a quartic selfinteraction potential $\lambda|\phi|^4$, and has been established through a combination of semianalytic studies of the linearized equations~\cite{Gleiser:1988rq, Gleiser:1988ih} and numerical simulations~\cite{Seidel:1990jh, Guzman:2009xre} of the fully nonlinear field equations in spherical symmetry. Regarding the excited states, there exist different studies in the literature with different conclusions. On the one hand, in Ref.~\cite{JETZER1990265} Jetzer argues that for the spherically symmetric excited states there also exists stable and unstable branches that are divided by the state of maximum mass, similarly to what happens for ground state configurations, although his analysis relies on a pulsation equation which is singular at the nodes. In Ref.~\cite{Lee:1988av}, Lee and Pang use different analytical arguments to conclude that, on the contrary, excited states are unstable, at least in absence of selfinteractions. Finally, based on numerical evolutions of the spherically symmetric Einstein-Klein-Gordon equations, Balakrishna {\it et al.}~\cite{Balakrishna:1997ej} have confirmed that excited configurations are unstable, even if quartic selfinteractions are considered. Full 3D numerical evolutions are currently available and lead to similar conclusions regarding the stability of boson stars under generic perturbations (for details, see e.g. Sec.~4 of~\cite{Liebling:2012fv} and references therein).

However, in a series of recent papers, Sanchis-Gual {\it et al.}~\cite{Sanchis-Gual:2021phr} and Brito {\it et al.}~\cite{Brito:2023fwr} have argued that if the selfinteraction is repulsive and strong enough, excited boson stars may be stable. Their conclusion is based on the numerical evolution of perturbed boson stars in the spherically symmetric sector of the fully nonlinear Einstein-Klein-Gordon theory with quartic selfinteractions. In particular, they find excited configurations in which no apparent instabilities are manifest during the time span of the evolution, indicating that these states are either stable or are unstable with a large time scale associated with the unstable modes. Of course, the restriction to spherical symmetry  leaves open the possibility that such excited states, although stable with respect to radial perturbations, suffer from instabilities with respect to generic perturbations.

The purpose of this paper is to shed new light on the stability problem of excited boson stars in the selfinteracting case. For this, we concentrate on the Newtonian limit of the theory, where the complications of relativistic effects are absent and which, as we show, allows for a systematic study of this problem in the linearized case. In the nonrelativistic limit the Einstein-Klein-Gordon system reduces to the Gross-Pitaevskii-Poisson~\cite{Zhang:2018slz, Guth:2014hsa, Dmitriev:2021utv} or the Schr\"odinger-Poisson~\cite{1999Nonli..12..201T, Moroz:1998dh, 2002math.ph...8045H} system, depending on whether or not the theory includes a selfinteraction term. The Gross-Pitaevskii-Poisson system is particularly relevant for the study of axion dark matter candidates (see for instance Refs.~\cite{Sikivie:2006ni, Hu:2000ke, Matos:2000ss, Arvanitaki:2009fg, Suarez:2013iw, Marsh:2015xka, Hui:2016ltb, Niemeyer:2019aqm, Urena-Lopez:2019kud, Ferreira:2020fam}), and it is the main target of this paper. In galaxies, and leaving the central region apart (which is baryon dominated), visible matter moves at velocities that are much smaller than the speed of light, signaling that 
it might be sufficient to describe dark matter haloes using Newtonian physics. In addition, since the axion potential is non-linear~\cite{GrillidiCortona:2015jxo} selfinteractions are indeed expected to play a relevant role.

In particular, our study is based on some analytic and numerical methods that have been previously developed to analyze the linear stability of the equilibrium configurations of the Schr\"odinger-Poisson system~\cite{Roque:2023sjl, Nambo:2023yut}.  When applied to the Gross-Pitaevskii-Poisson system, we obtain that spherically symmetric ground state configurations are stable if the selfinteraction is repulsive, in analogy to what happens in the non-selfinteracting case, and there exists a state of maximum mass that divides the stable and the unstable branch in the case of an attractive selfinteraction~\cite{2011PhRvD..84d3531C,2011PhRvD..84d3532C}. This is similar to what happens for relativistic non-selfinteracting boson stars, although it is important to stress that the existence of a maximum mass state is now determined by the attractive selfinteraction and not by relativistic effects. Furthermore, our analysis allows us to consider generic linear perturbations and is not restricted to the radial case.

Regarding the excited states, our study reveals that, even if they are, in general, unstable, there exist configurations belonging to the first excited state that remain stable under generic linear perturbations when the effects of a repulsive selfinteraction become significant. More specifically, for the solution space that we have explored in this paper, if the selfinteraction is attractive, as well as in absence of selfinteractions, spherically symmetric excited boson stars are unstable under radial perturbations and hence generically unstable. However, in case that the selfinteraction is repulsive, there exist excited configurations that remain stable under radial perturbations, at least for the first two excited states that we have considered. The existence of radially stable excited states is consistent with the results reported in Refs.~\cite{Sanchis-Gual:2021phr, Brito:2023fwr}, as will be discussed. Furthermore, we extend this analysis to consider generic perturbations that do not necessarily respect the spherical symmetry of the stationary states and conclude that only a small region in the solution space corresponding to the first excited states remains stable under generic perturbations.

In this paper we work in natural units where $c=\hbar=1$ and use the $(-,+,+,+)$ signature convention for the spacetime metric. For convenience we sometimes express Newton's constant $G$ in terms of the Planck mass, $M_{\textrm{Pl}} \equiv 1/\sqrt{G}$.

\section{Selfinteracting scalar fields}\label{II}

Our starting point is the Einstein-Klein-Gordon theory for a complex scalar field $\phi(t, \vec{x})$ of mass $m_0$ and quartic selfinteraction $\lambda |\phi|^4$. This theory is described in terms of the action
\begin{subequations}\label{eq.action}
\begin{align}\label{eq.action1}
S[g_{\mu\nu},\phi]=& \int d^4x\sqrt{-g}\left(\frac{1}{16\pi G}R+\mathcal{L}_M\right),
\end{align}
which consists on the Einstein-Hilbert term with matter sector
\begin{equation}\label{eq.action2}
\mathcal{L}_M=-\nabla_{\mu}\phi^*\nabla^{\mu}\phi-m_{0}^{2}\abs{\phi}^2-\lambda\abs{\phi}^4.
\end{equation}
\end{subequations}
As usual, $g$ is the determinant of the spacetime metric $g_{\mu\nu}$, $R$ is the Ricci scalar, and $\phi^*$ denotes the complex conjugate of $\phi$, with $\abs{\phi}^2$ its modulus square.
The coupling constant $\lambda$ is dimensionless and can take both signs, depending on whether the selfinteraction is repulsive ($\lambda>0$) or attractive ($\lambda<0$). In case that $\lambda$ vanishes we recover a theory with no selfinteractions, where the scalar field is only coupled to gravity. For illustrative purposes we have restricted ourselves to a quartic selfinteraction, although our conclusions are independent of the potential, as we clarify in the next subsection and the Appendix~\ref{app.generic.potential}.

\subsection{Nonrelativistic limit}\label{sec:nonrelativistic}

In the nonrelativistic regime we are interested in  it is convenient to write the spacetime line element in the form\footnote{Equation~(\ref{eq.metric.approx}), which is expressed in the Newtonian gauge, codifies only the scalar degrees of freedom of the gravitational field (i.e. for fixed $t$, the fields $\Phi(t,\vec{x})$ and $\Psi(t,\vec{x})$ transform as scalars under spatial rotations). Vector and tensor modes do not couple to nonrelativistic matter and we have not included them here for simplicity.} 
\begin{subequations}\label{eqs.weak.nonrel.decomposition}
\begin{equation}\label{eq.metric.approx}
 ds^2=-\left[1+2\Phi(t,\vec{x})\right]dt^2+\left[1-2\Psi(t,\vec{x})\right]\delta_{ij}dx^idx^j,
\end{equation}
and decompose the scalar field into
\begin{equation}
 \phi(t,\vec{x})= \frac{1}{\sqrt{2m_0}}e^{-im_0t}\psi(t,\vec{x}),
\end{equation}
\end{subequations}
where $\Phi(t,\vec{x})$ and $\Psi(t,\vec{x})$ are the gravitational potentials and $\psi(t,\vec{x})$ is the wave function. In the nonrelativistic limit the different quantities scale as $\partial_t \sim \epsilon^{1/2}\partial_i\sim \epsilon m_0$, $\Phi\sim\Psi\sim \epsilon$ and $\psi\sim \sqrt{M_{\textrm{Pl}}^2m_0}\epsilon$, with $\epsilon$ a small positive number.

Introducing the decomposition~(\ref{eqs.weak.nonrel.decomposition}) into the action~(\ref{eq.action}) and working to the lowest orders in $\epsilon$ we arrive at: 
\begin{align}
S[&\Phi,\Psi,\psi]=\int dt\int d^3x \left[\frac{1}{8\pi G}\Psi\Delta\left(2\Phi-\Psi\right)\right. \nonumber \\ 
&\left.+\psi^*\left(i\frac{\partial}{\partial t}+\frac{1}{2m_0}\Delta -\frac{\lambda}{4m_0^2}|\psi|^2 \right)\psi-m_0\Phi|\psi|^2\right],\label{eq.action.nonrel}
\end{align}
where $\Delta$ refers to the three-dimensional flat Laplace operator. In order of appearance, the first term of this equation describes the gravitational field, the second term the matter sector and the last one the interaction between the two. Note that there are no time derivatives of the gravitational potentials in the action of Eq.~(\ref{eq.action.nonrel}). This is a consequence of the fact that gravity is not dynamical in the nonrelativistic limit.

Equation~(\ref{eq.action.nonrel}) is cubic in the small parameter $\epsilon$, except for the selfinteraction term which contains four powers of $\epsilon$, which suggests that its effect is negligible for small field amplitudes. However, this term is multiplied by the coupling constant $\lambda$, which indicates that it starts to contribute when the amplitude of the field is sufficiently large, such that $\epsilon\sim m_0^2/(\lambda M_{\textrm{Pl}}^2)$. This allows us to introduce the  selfinteraction parameter
\begin{equation}\label{eq.Lambda}
 \Lambda:=\frac{|\lambda|M_{\textrm{Pl}}^2}{2\pi m_0^2},
\end{equation}
which is dimensionless and measures the ``strength" of the selfinteraction. Note that, given that $m_0$ is expected to be much smaller than the Planck mass, the parameter $\Lambda$ is naturally large. 
In Appendix~\ref{app.generic.potential} we argue that Eq.~(\ref{eq.action.nonrel}) is, in fact, valid for potentials $V(\phi)$ which are more general than the quartic selfinteraction one.

The variation of Eq.~(\ref{eq.action.nonrel}) with respect to the wave function $\psi$ results in the Gross-Pitaevskii equation~\cite{Gross1961, Pitaevskii1961}
\begin{subequations}\label{eqs.GPP}
\begin{equation}\label{eq.GPP}
 i\frac{\partial\psi}{\partial t} = -\frac{1}{2m_0}\Delta\psi \pm\frac{\pi\Lambda}{M_{\textrm{Pl}}^2}|\psi|^2\psi +m_0\mathcal{U}\psi,
\end{equation}
whereas the variation with respect to $\Psi$ yields (after integration by parts and discarding boundary terms) $\Delta(\Phi - \Psi) = 0$. Assuming that $\Phi$ and $\Psi$ vanish at infinity, this implies that $\Phi=\Psi$. Finally, variation of Eq.~(\ref{eq.action.nonrel}) with respect to $\Phi$ results in the Poisson equation
\begin{equation}\label{eq.Poisson}
 \Delta\mathcal{U}=4\pi G m_0|\psi|^2
\end{equation}
\end{subequations}
for the gravitational potential $\mathcal{U}:=\Phi=\Psi$. The $\pm$ signs in Eq.~(\ref{eq.GPP}) make reference to the repulsive ($+$) and attractive ($-$) cases. 
We will refer to the system of equations~(\ref{eqs.GPP}) as the Gross-Pitaevskii-Poisson system. Note that if the selfinteraction term vanishes, i.e. $\Lambda=0$, this reduces to the more familiar Schr\"odinger-Poisson system.

The nonrelativistic action, Eq.~(\ref{eq.action.nonrel}), is invariant under time translations, which means that the evolution of the system is not affected by shifts in the time parameter $t$. Associated to this symmetry we can define the total energy of the configuration,
\begin{align}
\mathcal{E} =&\int \left(\frac{1}{2 m_0}\abs{\nabla \psi}^2\pm\frac{\pi\Lambda}{2M_{\textrm{Pl}}^2}\abs{\psi}^4+\frac{1}{2}m_0\mathcal{U}|\psi|^2\right)d^3x,
\label{Eq:TotalEnergyNR}
\end{align}
which is conserved during the evolution. In addition, Eq.~(\ref{eq.action.nonrel}) is also invariant under continuous shifts in the phase of the wave function, $\psi\mapsto e^{i\alpha}\psi$, with $\alpha$ a real constant, which results in the conservation of the particle number
\begin{equation}\label{Eq:PartNumb}
N = \int |\psi|^2d^3x.
\end{equation}
Other conserved quantities associated with the Galilei group do exist; however, they will not be used in this article. For a systematic study of symmetries and conserved quantities for $\Lambda=0$, see Refs. \cite{oRpT06, Giulini_2011, Duval_2015}.

\subsection{Reformulation in terms of dimensionless quantities}

In this section we formulate the Gross-Pitaevskii-Poisson system in a more convenient form. To proceed, we introduce the dimensionless quantities
\begin{subequations}\label{eq.code.numbers1}
\begin{eqnarray}
&{\displaystyle \bar{t}:=\frac{2 m_{0} }{\Lambda}t, \quad \bar{x} := \frac{2 m_{0}}{\Lambda^{1/2}} x,}\\ \label{Eq:LambdaRescaling}
&{\displaystyle \bar{\mathcal{U}} := \frac{\Lambda}{2}\mathcal{U}, \quad \bar{\psi} := \left(\frac{ \pi\Lambda^2}{2 M_{\textrm{Pl}}^2m_0}\right)^{1/2}\psi.}
\end{eqnarray}
\end{subequations}	
In terms of these new variables the 
Gross-Pitaevskii~(\ref{eq.GPP}) and Poisson~(\ref{eq.Poisson}) equations simplify to
\begin{subequations}\label{eq.dimensionless.GPP.1}
\begin{align}
i\frac{\partial \psi}{\partial t} &=\left(-\Delta\pm|\psi|^2+\mathcal{U}\right)\psi,\\
\Delta \mathcal{U} &= |\psi|^2,
\end{align}
\end{subequations}
where we have omitted the bars in order to simplify the presentation (from now on we will denote dimensionfull quantities with the superscript {\it phys} whenever necessary).

Equivalently, Eqs.~(\ref{eq.dimensionless.GPP.1}) can be expressed as an integro-differential nonlinear equation
\begin{equation}\label{Eq:SPA_NLsystem}
	i\frac{\partial \psi}{\partial t} = \hat{{\mathcal{H}}}(\psi)\psi,
\end{equation}
where we have defined the dimensionless operator
\begin{equation}\label{eq.def.H}
   \hat{\mathcal{H}}(\psi) := -\triangle \pm |\psi|^2 + \triangle^{-1} (|\psi|^2),
\end{equation}
and where for a generic function $f(\vec{x})$ we have introduced
\begin{equation}
\Delta^{-1}(f)(\vec{x}) := -\frac{1}{4\pi}\int \frac{f(\vec{y})}{|\vec{x} - \vec{y}|}d^3 y.
\end{equation}
Note that in terms of the variables~(\ref{eq.code.numbers1}) the parameter $\Lambda$ disappears from our equations; hence, all possible values of the coupling constant $\lambda$ can be explored at the same time, implying that our results do not depend on the strength of the selfinteraction.

In terms of the dimensionless quantities the conserved energy functional (\ref{Eq:TotalEnergyNR}) $\mathcal{E}^{phys} [\psi^{phys}] = [M_{\textrm{Pl}}^2/(m_0\Lambda^{3/2}\pi)]\mathcal{E} [\psi]$ can be expressed in the form
\begin{equation}
   \mathcal{E}[\psi] = T[\psi] \pm F[n] - D[n, n], \quad n := |\psi|^2,
   \label{DimensionlessFuncE} 
\end{equation}
where the functionals $T$, $F$ and $D$ are defined by 
\begin{subequations}\label{Eqs:relations}
\begin{align}
    T[\psi] &:= \frac{1}{2}\int |\nabla \psi(\vec{x})|^2 d^3x,\\ 
    F[n] &:= \frac{1}{4} \int n(\vec{x})^2 d^3x,\\
    D[n, n] &:= \frac{1}{16\pi} \int \int \frac{n(\vec{x}) n(\vec{y})}{|\vec{x} - \vec{y}|} d^3y d^3x.
\end{align}
\end{subequations}
Furthermore, the first and second variations of the energy functional are given by (see Appendix~B of Ref.~\cite{Roque:2023sjl} for some details on a similar calculation)
\begin{subequations}\label{Eq:Variation}
\begin{align}
\label{Eq:FirstVariation}
\delta\mathcal{E} &= \Re( \hat{\mathcal{H}}(\psi)\psi,\delta\psi), \\
\label{Eq:SecondVariation}
\delta^2\mathcal{E} &= \Re(\hat{\mathcal{H}}(\psi)\psi,\delta^2 \psi)
 + (\hat{\mathcal{H}}(\psi)\delta\psi,\delta\psi)
 \pm 2F[\delta n]\\ \nonumber
 &- 2D[\delta n,\delta n],
\end{align}
\end{subequations}
with $\delta n := 2\Re(\psi^*\delta \psi)$ and $(\psi,\phi)=\int \psi^*\phi d^3x$ denoting the standard $L^2$-scalar product between $\psi$ and $\phi$, such that $(\psi,\psi) = N$. 

\section{Stationary states and their linear stability}

In this section we focus on scalar field  configurations of the form:
\begin{equation}\label{eq.harmonic}
 \psi(t,\vec{x})=e^{-i E t} \sigma^{(0)}(\vec{x}) ,
\end{equation}
where $\sigma^{(0)}$ is a real-valued function of $\vec{x}$ and $E$ a real constant.\footnote{Since the operator $\hat{\mathcal{H}}(\sigma^{(0)})$ is real, in the sense that $\hat{\mathcal{H}}(\sigma^{(0)})\psi^* = (\hat{\mathcal{H}}(\sigma^{(0)})\psi)^*$, there is no restriction in demanding that $\sigma^{(0)}$ is real-valued.} Introducing this ansatz into Eq.~(\ref{Eq:SPA_NLsystem}) we arrive at the nonlinear eigenvalue problem:
\begin{equation}\label{Eq:NLsystem}
E\sigma^{(0)} =\hat{\mathcal{H}}(\sigma^{(0)})\sigma^{(0)}.
\end{equation}
The integro-diferential equation~(\ref{Eq:NLsystem}) determines the stationary solutions to the Gross-Pitaevskii-Poisson system. 

The stability of these solutions is determined by the 
behavior of the small deviations about Eq.~(\ref{eq.harmonic}). To proceed, we linearize the integro-differential equation~(\ref{Eq:SPA_NLsystem}) following the procedure presented in Refs.~\cite{2002math.ph...8045H, Roque:2023sjl}. With this is mind, we propose the following ansatz for the wave function
\begin{equation}
\psi (t,\vec{x})= e^{-i E t}\left[\sigma^{(0)}(\vec{x}) 
 + \epsilon\sigma(t,\vec{x})+\mathcal{O}(\epsilon^2) \right],
\label{eq:ansatzPert}
\end{equation}
where $\epsilon$ is a small positive parameter. Here, $(E, \sigma^{(0)})$ is a solution of the nonlinear eigenvalue problem~(\ref{Eq:NLsystem}) and $\sigma$ is a complex-valued function depending on $(t,\vec{x})$ that describes the linear perturbation.

Introducing the ansatz (\ref{eq:ansatzPert}) into Eq.~(\ref{Eq:SPA_NLsystem}) one obtains, to linear order in $\epsilon$,
\begin{align}
i\frac{\partial \sigma}{\partial t} =&\left(\hat{\mathcal{H}}^{(0)}-E\right)\sigma+2\sigma^{(0)}\hat{K}\left[\sigma^{(0)}\Re(\sigma)\right],
\label{Ec:FirstOrd}
\end{align}
with the linear operators $\hat{\mathcal{H}}^{(0)} := \hat{\mathcal{H}}(\sigma^{(0)})$ and
\begin{equation}
    \hat{K}:=\pm 1 + \triangle^{-1}.
\label{eq:KDef}
\end{equation}
To separate the temporal from the spatial parts of $\sigma$ we use the mode ansatz (see also Sec.~5.2 in~\cite{10.5555/1941970} for details):
\begin{equation}
\sigma(t,\vec{x}) = \left[ A(\vec{x})+B(\vec{x}) \right]e^{\lambda t} + \left[A(\vec{x})-B(\vec{x})\right]^{*}e^{\lambda^* t},
\label{Ecpert}
\end{equation}
where $A$ and $B$ are complex-valued functions depending only on $\vec{x}$ and $\lambda$ is a complex constant (not to be confused with the selfinteraction coupling constant $\lambda$ of Eq.~(\ref{eq.action2})). Substituting Eq.~(\ref{Ecpert}) into Eq.~(\ref{Ec:FirstOrd}) and setting the coefficients in front of $e^{\lambda^* t}$ and $e^{\lambda t}$ to zero, we arrive at
\begin{subequations}
\label{SecOrd12lB}
 \begin{align}
i\lambda A=&\left(\hat{\mathcal{H}}^{(0)}-E\right)B, \label{SecOrd12blB}\\
i\lambda B=&\left(\hat{\mathcal{H}}^{(0)}-E\right) A+2\sigma^{(0)} \hat{K}\left[ \sigma^{(0)}A \right].
\label{SecOrd12alB}
\end{align}
\end{subequations}
This system constitutes a linear eigenvalue problem for the constant $\lambda$. Notice that linear instability is signaled by the existence of solutions with a positive real part $\lambda_R$ of $\lambda$. The lifetime of the unstable configurations is expected to be of the order of $t_{\textrm{life}}\sim 1/\lambda_R^{\textrm{max}}$, with $\lambda^{\textrm{max}}$ the eigenvalue with the largest real part.

\subsection{Basic properties of the stationary states}
\label{subsec.scalling}

Next, we present some basic properties satisfied by the solutions of Eq.~(\ref{Eq:NLsystem}), based on a simple scaling argument similar to the one used in Ref.~\cite{1964JMP.....5.1252D}. In Sect.~\ref{sect.stability.scalling} these properties will be used to shed light on the stability of stationary states in the attractive case.

To explain our scaling argument, recall that stationary states are critical points of the energy functional~(\ref{DimensionlessFuncE}), assuming that the particle number $N = (\psi,\psi)$ is fixed. To show this we may use Eqs.~(\ref{Eq:FirstVariation}) and~(\ref{Eq:NLsystem}), which imply that stationary states~(\ref{eq.harmonic}) satisfy $\delta\mathcal{E} = E\Re\left( \sigma^{(0)},\sigma \right)$. On the other hand, since $N$ is fixed $0 = \Re(\psi,\delta\psi) = \Re(\sigma^{(0)},\sigma)$, which implies that $\delta\mathcal{E} = 0$.

Now, let $\nu > 0$ be an arbitrary real and positive parameter and $\psi(t,\vec{x})$ a given wave function. Consider the rescaled function
\begin{equation}\label{eq.rescaling}
\psi_\nu(t,\vec{x}) := \nu^{3/2}\psi(t,\nu\vec{x}),
\end{equation}
which leaves the particle number invariant: $(\psi_\nu,\psi_\nu) = (\psi,\psi)$. 
Under this rescaling, the energy  functional~(\ref{DimensionlessFuncE}) transforms as
\begin{equation}
    \mathcal{E}[\psi_\nu] = \nu^2 T[\psi] \pm \nu^3 F[n] - \nu D[n,n],\label{Eq:Rescaling}
\end{equation}
and the first and second variations of $\mathcal{E}[\psi_\nu]$ at $\psi_{\nu=1} = \psi$ are
\begin{subequations}
\begin{align}
    \label{FirstDetivative}
    \left.\frac{d} {d\nu} \mathcal{E}[\psi_\nu] \right|_{\nu = 1} &= 2T[\psi] \pm 3F[n] - D[n,n], \\
    \label{SecondDerivative}
    \left.\frac{d^2} {d\nu^2} \mathcal{E}[\psi_\nu] \right|_{\nu = 1}  &= 2T[\psi] \pm 6F[n].
\end{align}
\end{subequations}
In particular, if $\psi$ is a stationary solution, the first variation is zero, which yields the relation
\begin{equation}\label{Eq:EnergFunctStat}
D[n,n] = 2T[\psi] \pm 3 F[n].
\end{equation}
This expression is valid for any stationary solution and allows one to eliminate $D[n,n]$ in the energy functional and compute the energy for such states solely in terms of $T[\psi]$ and $F[n]$ according to
\begin{equation}\label{Eq:EnerRelat}
\mathcal{E}[\psi] = -T[\psi] \mp 2F[n].
\end{equation}
This expression implies that the energy of the stationary states is always negative in the repulsive case. In the attractive case, it follows from Eq.~(\ref{Eq:Rescaling}) that $\mathcal{E}[\psi_\nu]$ is not bounded from below, since it can be made arbitrarily negative by choosing $\nu$ large. According to Eq.~(\ref{SecondDerivative}), the critical point at $\nu=1$ corresponds to a local minimum of $\mathcal{E}[\psi_\nu]$ if $T - 3F > 0$ and to a local maximum if $T - 3F < 0$. It follows from these observations that in the attractive case a stationary state $\psi$ cannot be a (local) minimum of the energy functional if $T < 3F$, that is, when the selfinteraction term dominates the kinetic term.

\subsection{Basic properties of the linearized equations}

We close this section by reviewing some properties of the solutions to the system of Eqs.~(\ref{SecOrd12lB}) that describe the behavior of linear perturbations. These properties and their derivation are completely analogous to the ones reported in~\cite{Nambo:2023yut}; hence, we only mention the most relevant ones.

First, note that there always exists the zero-mode solution $(\lambda, A, B) = (0, 0, \beta \sigma^{(0)})$, with $\beta$ an arbitrary complex constant. This corresponds to the trivial perturbation that rotates the real function $\sigma(\vec{x})$ into the complex plane, resulting in a new configuration $\psi(t, \vec{x})$ that is indistinguishable from~(\ref{eq.harmonic}). 

Second, the solutions to the Eqs.~(\ref{SecOrd12lB}) appear always in quadruples, that is, if $(\lambda, A, B)$ is a solution, then $(-\lambda, A, -B)$, $(\lambda^{*}, A^{*}, -B^{*})$ and $(-\lambda^{*}, A^{*}, B^{*})$ are also solutions. This implies that the eigenvalues $\lambda$ appear always in the form $\{\lambda, -\lambda, \lambda^*, -\lambda^*\}$. Linear stability requires that all eigenvalues $\lambda$ are purely imaginary. In the remaining of this work we shall count the number of unstable modes by the number of eigenvalues with distinct real part. Hence, when both the real and imaginary parts of $\lambda$ are different from zero, we count the pair of eigenvalues $\lambda$ and $\lambda^*$ as one unstable mode, although they may belong to two linearly independent eigenfunctions.

Third, from Eq.~(\ref{Eq:SecondVariation}) we obtain the following expression for the second variation of the energy functional,
\begin{equation}
\delta^2\mathcal{E} = (\delta\psi, [\hat{\mathcal{H}}^{(0)} - E]\delta \psi) \pm 2F[\delta n] - 2D[\delta n, \delta n],
\label{Eq:SecondVarStationary}
\end{equation}
which is quadratic in $\delta\psi$. Note that we have used the second variation of $(\psi,\psi) = N = const.$ to eliminate the term containing $\delta^2\psi$, and recall that $\delta n = 2\Re(\psi^*\delta\psi)$. The second variation of the energy functional $\delta^2\mathcal{E}$ is related to the linearized equations in the following way: multiplying Eq.~(\ref{SecOrd12alB}) by $A^{*}$ and Eq.~(\ref{SecOrd12blB}) by $B^{*}$ and integrating one obtains
\begin{subequations}
\label{Eq:ScalarProductAB}
\begin{align}
\label{Eq:ProductAB}
    i\lambda \left(A, B\right) &= \delta^2\mathcal{E}[A_R] + \delta^2\mathcal{E}[A_I],\\
    \label{Eq:ProductBA}
    i\lambda \left(B, A\right) &= \left(B, [\hat{\mathcal{H}}^{(0)} - E]B\right),
\end{align}
\end{subequations}
where $A_R, A_I$ denotes the real and imaginary parts of $A$, respectively, and $\delta^2\mathcal{E}[A_R]$ is given by Eq.~(\ref{Eq:SecondVarStationary}) replacing $\delta\psi$ with $A_R$. The right-hand sides of the previous equations are real; thus,
\begin{equation}
\label{Eq:ABReal}
    -\lambda^2|\left(A,B\right)|^2 \in \mathbb{R}.
\end{equation}
Finally, if $\lambda$ is real one can choose $A$ real and $B$ purely imaginary. Eliminating $i\lambda A$ on the left-hand side of Eq.~(\ref{Eq:ProductAB}) and using Eq.~(\ref{SecOrd12blB}) one gets
\begin{equation}
\label{Eq:Contradiction}
    -\left(B, [\hat{\mathcal{H}}^{(0)}-E] B \right) = \delta^2\mathcal{E}[A].
\end{equation}

Equations~(\ref{Eq:ScalarProductAB}), (\ref{Eq:ABReal}) and~(\ref{Eq:Contradiction}) are useful to rule out the existence of certain unstable modes. For example, consider a stationary state for which $E$ is the ground state energy of the Schr\"odinger operator $\hat{\mathcal{H}}^{(0)}$. Then, $\left(B, [\hat{\mathcal{H}}^{(0)}-E] B \right) \geq 0$ with equality if and only if $B$ lies in the kernel of $\hat{\mathcal{H}}^{(0)}-E$. This immediately excludes the existence of modes with $\lambda_R\neq 0$ and $\lambda_I\neq 0$ since in this case Eq.~(\ref{Eq:ABReal}) would imply that $(A,B) = 0$ and then one could infer that $A=B=0$ using Eqs.~(\ref{Eq:ProductBA}) and~(\ref{SecOrd12lB}). If, in addition, this state is a local minimum of the energy functional, such that $\delta^2\mathcal{E}$ is positive or zero, then unstable modes with $\lambda_I = 0$ are also ruled out since in this case Eqs.~(\ref{Eq:Contradiction}) and~(\ref{SecOrd12lB}) imply that $A=B=0$. These arguments show that stationary ground state configurations can have only real or purely imaginary  $\lambda$ and that real eigenvalues are excluded if these configurations are a local minimum of the energy functional $\mathcal{E}$.

Similar arguments will be applied to exclude other type of unstable modes in the next section.

\section{Stationary and spherically symmetric solutions}\label{Sec:IV}

\begin{figure*}
	\centering	
    \includegraphics[width=18.cm]{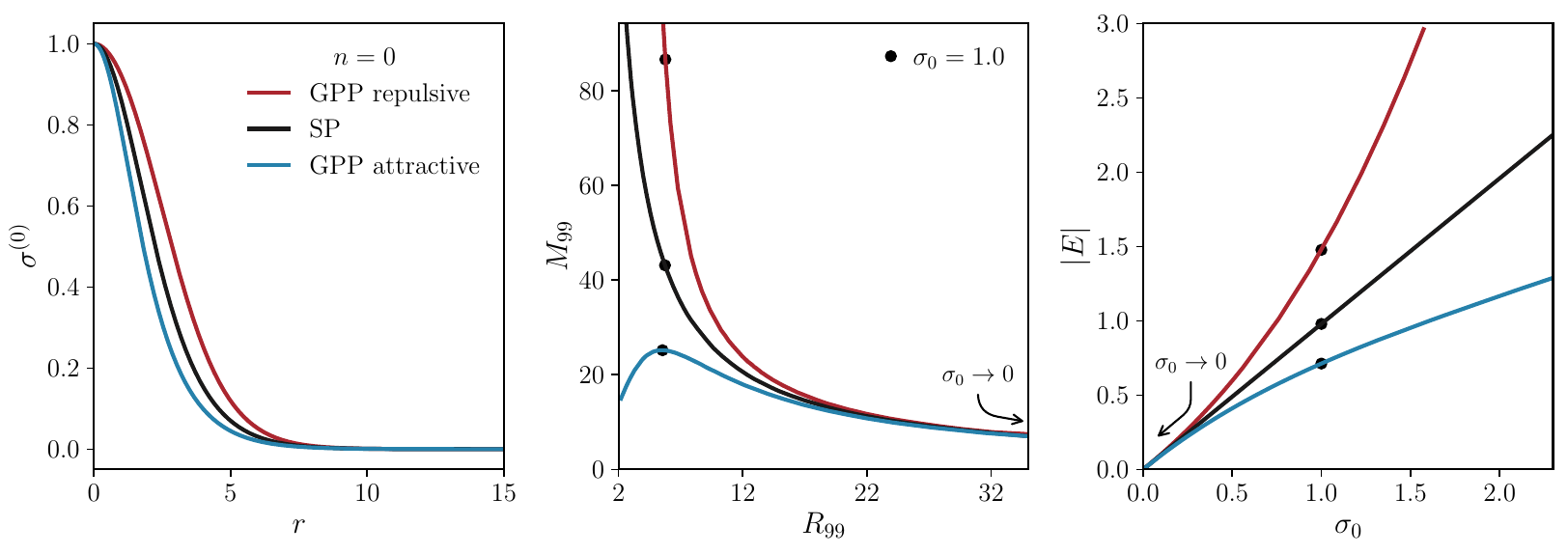}
\caption{{\bf Ground state configurations:} Stationary and spherically symmetric solutions of the Gross-Pitaevskii-Poisson system with no nodes ($n=0$). Red (blue) lines correspond to the repulsive (attractive) case, and we have included the solutions to the Schr\"odinger-Poisson system  (black lines) for reference.{\it (Left panel)} The profile of the wave function $\sigma^{(0)}(r)$ for a central value of unity, $\sigma_0=1$. {\it (Center panel)} The effective mass of the configurations $M_{99}$ as a function of the effective radius $R_{99}$. {\it (Right panel)} The magnitude of the energy eigenvalue $|E|$ as a function of the central amplitude $\sigma_0$. The dots in the last two panels correspond to the configurations of unit amplitude. For $\sigma_0\to 0$ the effects of the selfinteractions become negligible.}\label{FigFondn0}
\end{figure*}

\begin{figure*}
	\centering	
    \includegraphics[width=18.cm]{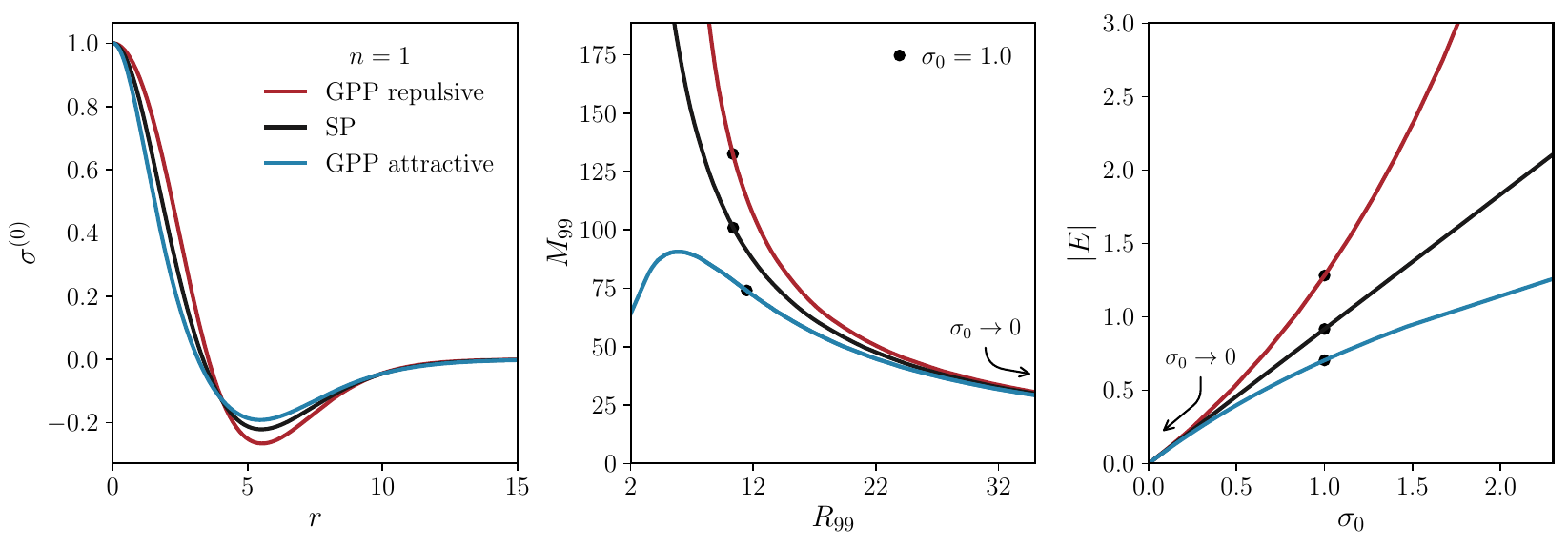}
	\caption{{\bf First excited state configurations:} Same as in Fig.~\ref{FigFondn0} but for the stationary and spherically symmetric solutions of the Gross-Pitaevskii-Poisson system with one node ($n=1$).}\label{FigFondn1}
\end{figure*}

\begin{figure*}
	\centering	
    \includegraphics[width=18.cm]{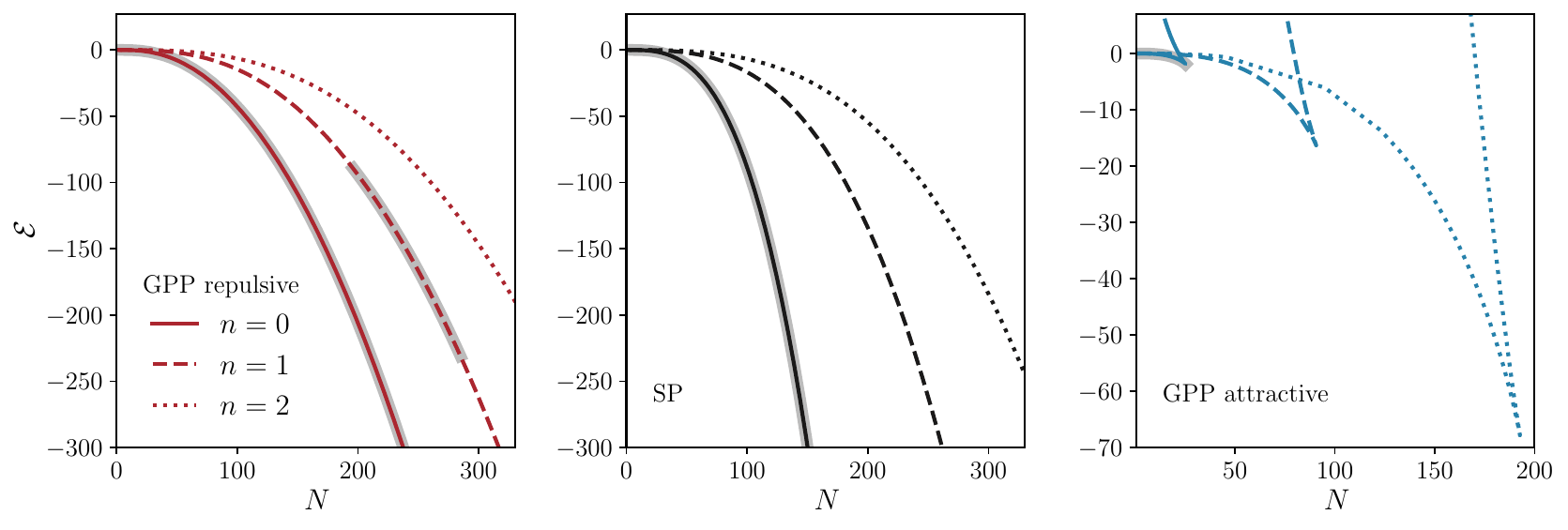}
    \caption{{\bf The energy $\mathcal{E}$ of the states as a function of their number of particles $N$:} Relationship between the energy functional~(\ref{Eq:EnerRelat}) and the particle number $N$ for states with zero, one and two nodes in the repulsive, non-selfinteracting and attractive cases. Note that in the attractive case the number of nodes $n$ does not necessarily label the number of the excited state. The shaded regions represent the stability bands with respect to generic linear perturbations discussed in Sec.~\ref{Sec:numerical.study}.}\label{fig.energNump}
\end{figure*}

For the remaining of this article we focus on stationary solutions which are spherically symmetric, i.e. $\sigma^{(0)}(\vec{x})=\sigma^{(0)}(r)$. In the next subsection we discuss the background (unperturbed) solutions, whereas some  properties of the linearized perturbations
are discussed in subsequent subsections. A detailed study of the linear stability based on a numerical analysis is presented in Sec.~\ref{Sec:numerical.study}.

\subsection{Background configurations}

Instead of solving directly the integro-differential equation~(\ref{Eq:NLsystem}), for a numerical analysis it is more convenient to work with the original Gross-Pitaevskii-Poisson system~(\ref{eq.dimensionless.GPP.1}). Introducing the harmonic ansatz~\ref{eq.harmonic}) into Eqs.~(\ref{eq.dimensionless.GPP.1}) and defining the shifted potential $u^{(0)}(r):= E - \mathcal{U}(r)$, the dimensionless Gross-Pitaevskii-Poisson equations take the form
\begin{subequations}\label{eq.dimensionless.GPP.3}
\begin{align}
\Delta_s \sigma^{(0)}&= \left(\pm \sigma^{(0)2}-u^{(0)}\right) \sigma^{(0)},\label{SEc2.2.2.1}\\
\Delta_s u^{(0)}&= - \sigma^{(0)2},
\end{align}
\end{subequations}
where $\Delta_s:=\frac{1}{r}\frac{d^2}{dr^2}r$ denotes the radial Laplace operator.

The system of equations~(\ref{eq.dimensionless.GPP.3}) must be solved for some appropriate boundary conditions. Regularity at the origin demands $\sigma^{(0)}(r=0)=\sigma_{0}, \sigma^{(0)\prime}(r=0)=0, u^{(0)}(r=0)=u_0$, and $u^{(0)\prime}(r=0)=0$. (Here and in the following the prime refers to derivation with respect to $r$.) Then, given $u_0$, the central amplitude of the field $\sigma_{0}$ is fine-tuned using a numerical shooting methodology based on the condition $\lim\limits_{r\to\infty}\sigma(r)=0$ at spatial infinity, which is required for the solution to be localized.

The system~(\ref{eq.dimensionless.GPP.3}) is solved numerically using an adaptive explicit 5(4)-order Runge-Kutta routine~\cite{2020SciPy-NMeth, DORMAND198019, Lawrence1986SomePR}, where we rewrite the equations as a first-order system for the fields $(\sigma^{(0)}, u^{(0)})$. For the shooting, we use a methodology similar to the one described in~\cite{Moroz:1998dh}, based on bisection. 

Some representative solutions of the stationary and spherically symmetric Gross-Pitaevskii-Poisson system are shown in Figs.~\ref{FigFondn0} and~\ref{FigFondn1} for the ground ($n=0$) and the first excited ($n=1$) states, respectively.

In these figures, the mass of the configurations $M^{phys}$ has been calculated as their particle number Eq.~(\ref{Eq:PartNumb}) times the mass $m_0$ of the individual particles, which leads to $M^{phys} = [M_{\textrm{Pl}}^2/(4\pi m_0 \Lambda^{1/2})]M$, where $M:=4\pi\int_0^{\infty} [\sigma^{(0)}(r)]^2 r^2dr$ is the dimensionless mass. Note that, formally, the radius of a boson star extends to infinity, and for that reason we have defined the effective radius $R_{99}$ as the one encompassing $99$\% of the total mass of the configuration, $M_{99}$. On the other hand, the energy eigenvalue $E^{phys}=[2m_0/\Lambda]E$ associated with the state $\sigma^{(0)}(r)$  can be computed from the formula (see Appendix~C in Ref.~\cite{Roque:2023sjl} for more details)
\begin{align}
E=u_0-\int_{0}^{\infty} [\sigma^{(0)}(r)]^2 r dr.
\end{align}

As expected, the profile of the configurations expands (shrinks) if the selfinteraction is repulsive (attractive), as can be appreciated in the left panels of Figs.~\ref{FigFondn0} and~\ref{FigFondn1}. Furthermore, in the repulsive case the configurations become more compact as we increase the value of the central amplitude, as also occurs for the stationary and spherically symmetric solutions of the Schr\"odinger-Poisson system, wheres in the case in which the selfinteraction is attractive there exists a state of maximum mass corresponding to a central amplitude $\sigma_0^{M_{99}^{\textrm{max}}}$; see the central panel of Figs.~\ref{FigFondn0} and~\ref{FigFondn1} and Refs.~\cite{2011PhRvD..84d3531C,2011PhRvD..84d3532C,Chavanis:2022fvh}. Finally, it is also interesting to stress that in the limit $\sigma_0\to0$ the attractive and the repulsive configurations approach the solutions to the Schr\"odinger-Poisson system, see the central and right panels of the same figures. This implies that the effects of the selfinteractions become negligible in the limit of low amplitudes, as is also apparent from the Gross-Pitaevskii-Poisson system~(\ref{eqs.GPP}).

In Fig.~\ref{fig.energNump} we plot the total energy $\mathcal{E}$ of the configurations as a function of the total number of particles $N$ for the ground and first two excited states. Note that the energy functional~(\ref{Eq:TotalEnergyNR}) consists of three terms: the first of them is a consequence of the ``quantum pressure'' and is positive definite. The second one is associated with the selfinteraction and can be positive or negative, depending on whether the selfinteraction is repulsive or attractive, respectively. Finally, the last term is associated with the gravitational interaction which is attractive and negative definite. It is interesting to note that in the repulsive case, as well as if there are no selfinteractions, the total energy is negative definite and its magnitude increases with the number of particles. Furthermore, for a given value of $N$, the energy also increases with the number of nodes $n$ in the configuration. This allows us to label the number of the excited states with $n$, as is usually done. However, this behavior changes for the case of an attractive selfinteraction, where, for a given $n$, there is a state of minimum energy (which coincides with the state of maximum mass), and from this point onward the energy increases up to positive values.

\subsection{Stability properties based on a scaling argument (attractive case)}\label{sect.stability.scalling}

\begin{figure*}
	\centering	
    \includegraphics[width=18.cm]{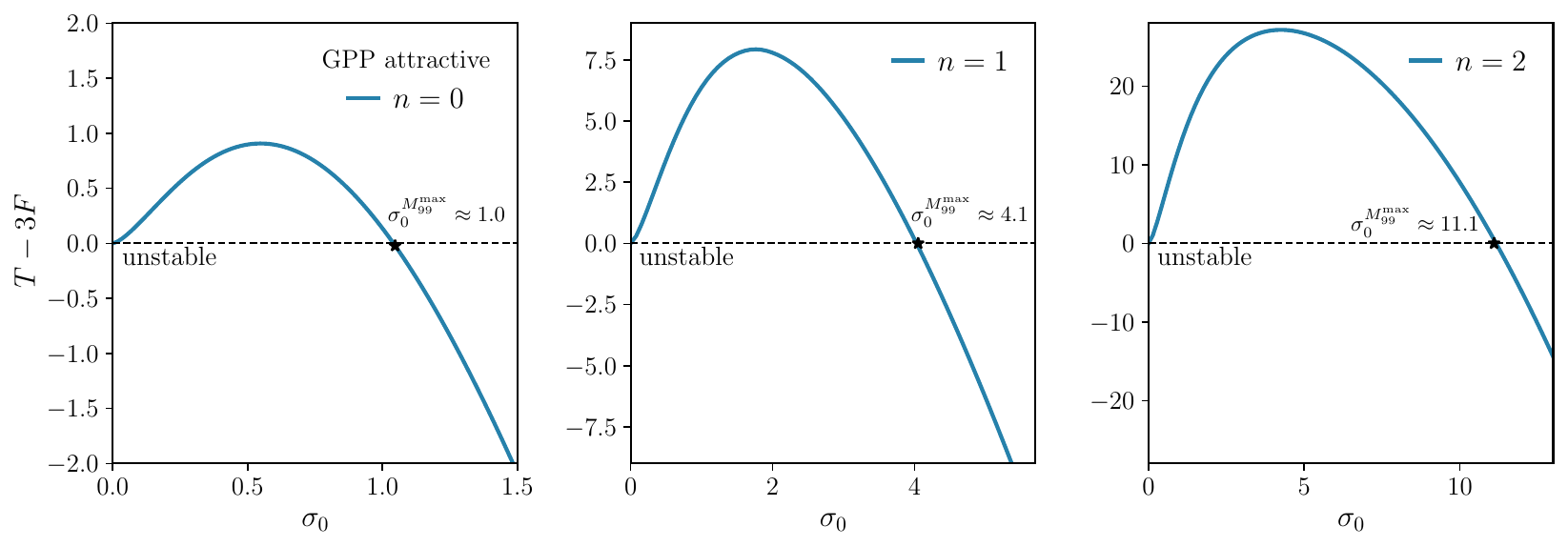}
	\caption{{\bf The combination $T-3F$ as a function of the central amplitude $\sigma_0$ for different values of the node number $n$ in the attractive case:} Configurations with a negative value of $T-3F$ correspond to stationary states that represent critical points of the energy functional which cannot be local minima, and hence they are expected to be unstable. The stars denote the maximum mass configurations at $\sigma_0= \sigma_0^{M^{\textrm{max}}_{99}}$; see also Fig.~\ref{fig.EnevsNu} which complements these plots.}\label{fig.second.variation}
\end{figure*}

\begin{figure*}
	\centering	
    \includegraphics[width=18.cm]{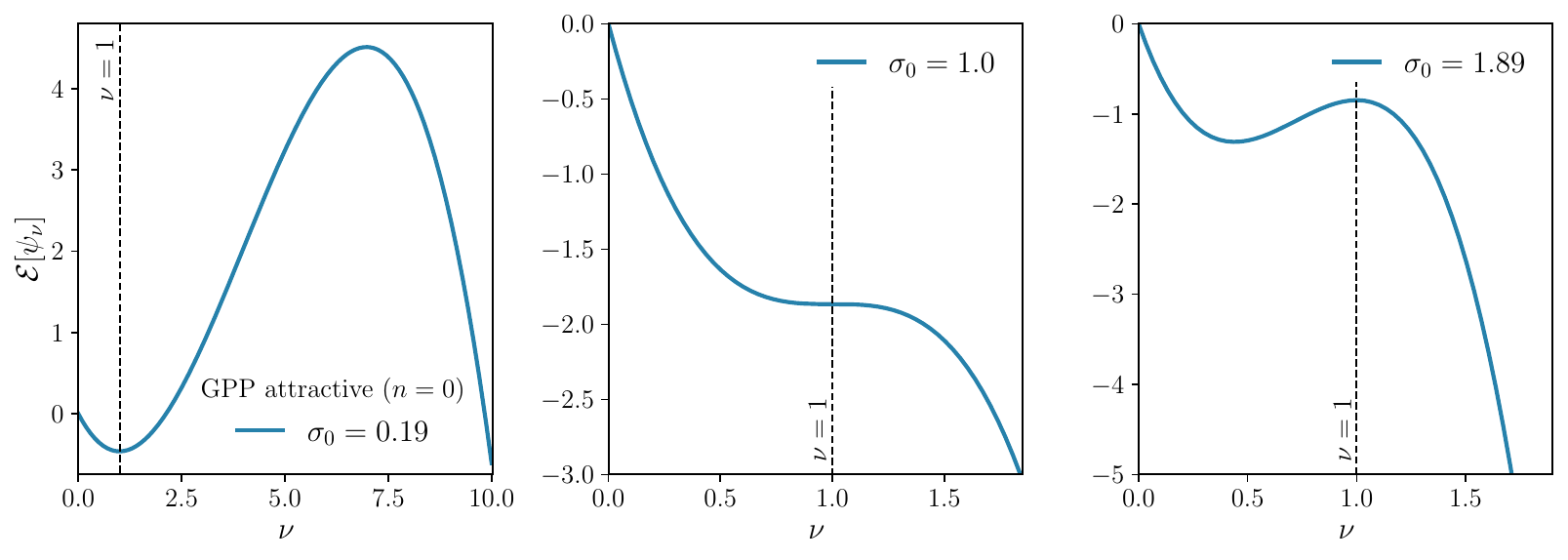}
	\caption{{\bf Scaling the energy functional as $\psi_\nu(t,\vec{x})=\nu^{3/2}\psi(t,\nu\vec{x})$ for attractive ground state configurations:} The energy functional $\mathcal{E}[\psi_\nu]$ as a function of $\nu$ for three ground state configurations. {\it (Left panel)} Stationary state with $\sigma_0 = 0.19<\sigma_0^{M_{99}^{\textrm{max}}}$ giving $T = 0.558$ and $F = 0.047$. This state could be a local (but not a global) minimum of the energy functional and hence could be stable. {\it (Center panel)} Stationary state with $\sigma_0=1.0=\sigma_0^{M_{99}^{\textrm{max}}}$ giving $T = 5.641$ and $F = 1.887$. This state resembles a saddle point of the energy functional. {\it (Right panel)} Stationary state with $\sigma_0 = 1.89$ giving $T = 11.411$ and $F = 5.282$. This state cannot be a local minimum of the energy functional and is expected to be unstable.}\label{fig.EnevsNu}
\end{figure*}

Recall from Sec.~\ref{subsec.scalling} that in the attractive case, a stationary configuration cannot be a local minimum of the energy functional if $T - 3F < 0$. When the central value of the scalar field $\sigma_0$ is small, the configurations resemble the ones of the Schr\"odinger-Poisson system, and hence the selfinteraction term $F$ is negligible, such that $T > 3F$ is expected to hold. In contrast, as  $\sigma_0$ grows, the influence of the selfinteraction becomes more pronounced (see Figs.~\ref{FigFondn0} and~\ref{FigFondn1}) and one expects $3F$ to surpass the value of $T$ after some critical value of $\sigma_0$.

The plots in Fig.~\ref{fig.second.variation} show that these expectations are indeed fulfilled. Interestingly, the critical value of $\sigma_0$ for which $T - 3F = 0$ seems to coincide with the maximum of the mass in the center panels of Figs.~\ref{FigFondn0} and~\ref{FigFondn1}. As will be discussed in the next section, when this critical value is surpassed, the number of unstable modes grows by one.

For completeness, Fig.~\ref{fig.EnevsNu} shows the behaviour of the energy functional under the rescaling~(\ref{eq.rescaling})  and illustrates that stationary states with positive (negative) values of $T - 3F$ represent a local minimum (maximum) with respect to this particular variation.

\subsection{Decomposition of the linearized system into spherical harmonics}

Since the background solution is spherically symmetric, the linearized equations can be decoupled into a family of purely radial systems by expanding the perturbations in terms of spherical harmonics $Y^{LM}$. In particular, the field $A(\vec{x})$ can be expanded as:
\begin{equation}
A(\vec{x}) = \sum\limits_{LM} A_{LM}(r) Y^{LM}(\vartheta,\varphi),
\end{equation}
and similarly for $B(\vec{x})$. This reduces Eqs.~(\ref{SecOrd12lB}) to the following system:
\begin{subequations}
\label{SecOrd12lBLM}
 \begin{eqnarray}
i\lambda A_{LM} &=& \left( \hat{\mathcal{H}}_L^{(0)} - E \right)B_{LM}, \label{SecOrd12blBLM}\\
i\lambda B_{LM} &=&\left(\hat{\mathcal{H}}_L^{(0)} - E \right) A_{LM} + 2\sigma^{(0)} \hat{K}_L\left[ \sigma^{(0)}A_{LM} \right],\qquad\;
\label{SecOrd12alBLM}
\end{eqnarray}
\end{subequations}
with the operators $\mathcal{\hat H}_L^{(0)}$ and $\hat{K}_L$ defined by
\begin{subequations}
\begin{align}
\hat{\mathcal{H}}_L^{(0)}&:=-\triangle_L 
 \pm \sigma^{(0)2} + \triangle_s^{-1}(\sigma^{(0)2}),\\
\hat{K}_L &:=\pm 1+\triangle_L^{-1},
\label{eq:H0LDef}
\end{align}
\end{subequations}
where $\triangle_L := \triangle_s - L(L+1)/r^2$ and
\begin{equation}
\triangle_L^{-1}(f)(r) := -\frac{1}{2L+1}\int\limits_0^\infty \frac{r_<^L}{r_>^{L+1}} f(\tilde{r})\tilde{r}^2 d\tilde{r},
\label{Eq:LapJInv}
\end{equation}
with $r_<:=\min\{ r,\tilde{r} \}$ and $r_>:=\max\{ r,\tilde{r} \}$.
Note, in particular, that $\Delta_s=\Delta_{L=0}$.

In the next section this system will be solved numerically for $L=0,1,2,\ldots$. Before doing that, however, we prove that no unstable modes are present if $L$ is large enough, which, in principle, reduces the numerical analysis to a finite number of $L$ values. 

\subsection{Non-existence of unstable modes for sufficiently large \texorpdfstring{$L$}{L}}\label{Sec:IVB}

In this subsection we show that if the orbital angular momentum $L$ of the perturbation is large enough, then there are no unstable modes. The arguments below generalize the ones in Sec.~IV.C of Ref.~\cite{Nambo:2023yut} to include the selfinteraction term.

Using the estimate (E8) for $D[\delta n,\delta n]$ in Appendix~E of Ref.~\cite{Nambo:2023yut} one obtains from Eq.~(\ref{Eq:SecondVarStationary}) the inequality\footnote{Note that the left-hand side of Eq.~(E8) in~\cite{Nambo:2023yut} should read $D[\delta n,\delta n]$ instead of $D[\delta u, \delta u]$.}
\begin{align}
\label{EstimateSecondVar}
   \delta^2\mathcal{E} &\geq \frac{1}{2}(\nabla \delta\psi, \nabla \delta\psi) + (\delta\psi, [\pm\sigma^{(0)2} + U_0 - E]\delta\psi)\\ \nonumber
   &- C_1(\delta\psi/f,\delta\psi/f) \pm 2F[\delta n], 
\end{align}
where $U_0 := \triangle^{-1}(|\sigma^{(0)}|^2)$, $C_1 > 0$ is a positive constant and $f(\vec{x}) := \sqrt{1 + |\vec{x}|^2}$. In the repulsive case, the term $+2F[\delta n]$ on the right-hand side of Eq.~(\ref{EstimateSecondVar}) is positive and hence can be discarded without altering the inequality. In the attractive case, we use the estimate $|\delta n|^2 \leq 4|\psi|^2 |\delta\psi|^2$, which implies
\begin{equation}
F[\delta n] \leq \max\limits_{\vec{x}\in\mathbb{R}^3} \left[ (1 + |\vec{x}|^2) |\psi(\vec{x})|^2 \right]
(\delta\psi/f,\delta\psi/f).
\end{equation}
Since $|\psi|$ is smooth and decays exponentially to zero as $|\vec{x}|\to \infty$,  the maximum is finite. Therefore, the negative term $-2F[n]$ on the right-hand side of Eq.~(\ref{EstimateSecondVar}) can be removed after making the constant $C_1$ larger.

Based on these results, we can prove that $\delta^2\mathcal{E}$ is positive definite for sufficiently large values of the orbital angular momentum $L$. To this purpose we expand $\delta\psi$ in spherical harmonics
\begin{equation}
    \delta\psi = \sum_{LM} h_{LM}(r) Y^{LM}(\vartheta, \varphi).
\end{equation}
Substituting this in expression (\ref{EstimateSecondVar}) and discarding the terms which involve $|h_{LM}'(r)|^2$  one gets 
\begin{align}
\label{EstimateSecondVarL}
    \delta^2\mathcal{E} &\geq \sum_{LM} \left\{\int_{0}^{\infty} |h_{LM}(r)|^2 \left[\frac{L(L + 1)}{2} - C_1\right] dr\right.\\ \nonumber
    &+ \left. \int_{0}^{\infty} |h_{LM}(r)|^2 [\pm\sigma^{(0)}(r)^2 + U_0(r) - E]r^2 dr\right\}.
\end{align}
First we look at the integrand in the second line and observe that the function $g(r) := [\pm\sigma^{(0)}(r)^2 + U_0(r) - E]r^2$ satisfies $g(0) = 0$ since $\sigma^{(0)}$ and $U_0$ are regular at the center, whereas $g(r)$ is positive for large enough $r$. This implies (since $U_0$ is continuous) that $g(r)$ has a minimum, that is, $g(r) \geq C_2$ for all $r \geq 0$ for some negative constant $C_2$. Using these properties, the estimate (\ref{EstimateSecondVarL}) yields
\begin{equation}
   \delta^2\mathcal{E} \geq \sum_{LM}\int_{0}^{\infty} |h_{LM}(r)|^2 \left[\frac{L(L + 1)}{2} - C_1 + C_2\right] dr.
\end{equation}
Therefore, $\delta^2 \mathcal{E}$ is positive definite if $h_{LM}$ is identically zero for all $L$ with $L(L + 1)/2 - C_1 + C_2 \leq 0$. This proves that the second variation of the energy functional is positive definite on the subspace of perturbations with large enough values of $L$. 

Now we can prove the non-existence of unstable modes for large $L$. We proceed by contradiction. If $\lambda$ has real and imaginary parts different from zero, then Eq.~(\ref{Eq:ABReal}) yields $(A, B) = 0$, and in this case Eq.~(\ref{Eq:ScalarProductAB}) and the positivity of $\delta^2\mathcal{E}$ imply that $A = B = 0$. On the other hand, if $\lambda$ is real, Eq.~(\ref{Eq:Contradiction}) leads to a contradiction since the right-hand side is positive whereas the left-hand side is negative for sufficiently large values of $L$.

\section{Numerical study of the linearized system}
\label{Sec:numerical.study}

To find the eigenvalues $\lambda$ of the system~(\ref{SecOrd12lBLM}) we use the methodology described in Refs.~\cite{Roque:2023sjl, Nambo:2023yut}, for which it is convenient to write the system of equations in a more appropriate form. For this, we define the new rescaled functions $a_L(r):=rA_{LM}(r)$ and $b_L(r):= rB_{LM}(r)$, in terms of which the system~(\ref{SecOrd12lBLM}) reduces to
\begin{subequations}\label{Ec2.2.2.2}
\begin{align}
&b_{L}'' \mp |\sigma^{(0)}|^{2} b_{L} + U^{\text{eff}}_{L}\,b_{L}=-i\lambda a_{L}, \\
&a_{L}'' + U^{\text{eff}}_{L}a_{L}\mp 3 |\sigma^{(0)}|^{2} a_{L}\\
&\qquad -2\alpha \sigma^{(0)}\left(\frac{d^2}{dr^2}-\frac{L(L+1)}{r^2}\right)^{-1}\left[\sigma^{(0)} a_{L}\right] =-i\lambda b_{L}.\nonumber\label{Eq48b}
\end{align}
\end{subequations}
Here we have introduced the effective potential $U^{\text{eff}}_{L}(r) := u^{(0)}(r) - L(L+1)/r^2$, and the operator $\left(d^2/dr^2-L(L+1)/r^2\right)^{-1} = r\triangle_L^{-1} r^{-1}$ denotes the inverse of $r \triangle_{L}(r^{-1})$ with homogeneous Dirichlet boundary conditions at $r=0$ and $r\to\infty$. 
Given that the system~(\ref{SecOrd12lBLM}) is independent of the magnetic quantum number $M$, we have omitted this label in Eqs.~(\ref{Ec2.2.2.2}).

These equations must be solved for some appropriate boundary conditions. To determine these conditions, one can study (heuristically) the dominant terms near the origin and at spatial infinity. Near $r=0$ the equations are dominated by the centrifugal term $L(L+1)/r^2$ of the effective potential $U^{\text{eff}}_{L}$, which means that regular solutions at the center must behave as $\left(a_{L}, b_{L})\sim(r^{L+1}, r^{L+1}\right)$. This leads to the following boundary conditions at the origin:
\begin{subequations}
\label{Eq:Dirichlet}
\begin{align}
    a_{L}(r=0)=0,\quad b_{L}(r=0)=0.
\end{align}
On the other hand, in the asymptotic region, the radial profile of the background field $\sigma^{(0)}(r)$ decays exponentially to zero and $u^{(0)}(r)$ approaches $E$. This implies that the fields $(a_{L}, b_{L})$ must vanish at infinity, 
\begin{align}
\lim\limits_{r\to\infty} a_{L}(r)=0,\qquad \lim\limits_{r\to\infty} b_{L}(r)=0,
\end{align} 
\end{subequations}
in order to have solutions with finite total energy.

To numerically solve the system~(\ref{Ec2.2.2.2}) with the Dirichlet boundary conditions~(\ref{Eq:Dirichlet}) we employ the following approach: first, the background profiles $\sigma^{(0)}(r)$ and $u^{(0)}(r)$ are computed following the method described in Sec.~\ref{Sec:IV}, and they are extended to large values of $r$ using the procedure described in Sec. III.A of~\cite{Roque:2023sjl}. Second, these background profiles, as well as the perturbed fields $a_{L}, b_{L}$ and the different operators, e.g. the derivative and its inverse, are discretized in terms of Chebyshev polynomials using a standard spectral method (see e.g. Ref.~\cite{trefethen2000spectral}), which leads to a finite-dimensional eigenvalue problem. For details of the numerical discretization procedure we refer the reader to Sec. IV of Ref.~\cite{Roque:2023sjl}.

The discrete version of the system~(\ref{Ec2.2.2.2}) can be writen as:
\begin{widetext}
\begin{gather}\label{Syst}
 \begin{pmatrix} \mathsf{0} & \mathbb{\tilde{D}}_{\mathsf{N}}^{2}\mp \Sigma_0^2+\mathsf{U}^{\text{eff}}_{L} \\ \mathbb{\tilde{D}}_{\mathsf{N}}^{2}\mp 3\Sigma_0^2+\mathsf{U}^{\text{eff}}_{L}-2\alpha\Sigma_{0}\big(\mathbb{\tilde{D}}_{\mathsf{N}}^{2}-\mathbb{L}\big)^{-1} \Sigma_0 & \mathsf{0}\end{pmatrix}\begin{pmatrix} \mathsf{a}_{L} \\ \mathsf{b}_{L}\end{pmatrix}
 =
 -i\lambda
   \begin{pmatrix} \mathsf{a}_{L} \\ \mathsf{b}_{L}\end{pmatrix},
\end{gather}
\end{widetext}
where here $\mathsf{0}$ represents the $(\mathsf{N}-1)\times (\mathsf{N}-1)$ zero matrix, with $\mathsf{N}$ the number of Chebyshev points distributed as $x_j=\cos{(j\pi/\mathsf{N})}, j=0,1,\dots,\mathsf{N}$. The quantities
\begin{subequations}
\begin{align}
\Sigma_0&:=\textbf{diag}\bigg(\sigma^{(0)}(x_1),\sigma^{(0)}(x_2),\dots,\sigma^{(0)}(x_{\mathsf{N}-1})\bigg),\nonumber\\
\mathsf{U}^{\text{eff}}_{L}&:=\textbf{diag}\bigg(U^{\text{eff}}_{L}(x_1), U^{\text{eff}}_{L}(x_2),\dots, U^{\text{eff}}_{L}(x_{\mathsf{N}-1})\bigg),\nonumber\\
\mathbb{L}&:=\textbf{diag}\bigg(\frac{L(L+1)}{x^{2}_1}, \frac{L(L+1)}{x^{2}_2},\dots,\frac{L(L+1)}{x^{2}_{\mathsf{N}-1}}\bigg),\nonumber
\end{align}
\end{subequations}
are the discrete representation of the background quantities and centrifugal term, respectively, and the discrete operator $\mathbb{\tilde{D}}_{\mathsf{N}}$ is defined in~\cite{Roque:2023sjl}. The vector
\begin{align}
\begin{pmatrix} \mathsf{a}_{L} \\ \mathsf{b}_{L}\end{pmatrix}
:=\bigg(a_{L}(x_1), \;\dots\; ,a_{L}(x_{\mathsf{N}-1}), b_{L}(x_1),\dots, b_{L}(x_{\mathsf{N}-1})\bigg)^{T}\nonumber
\end{align}
represents the eigenfields $r(A_{LM}, B_{LM})^T$. We solve the discrete eigenvalue problem~(\ref{Syst}) using the SciPy library~\cite{2020SciPy-NMeth}. Our code is publicly available in~\cite{Roque_On_the_radial_2023}. 

In the following subsections we present the results of the eigenvalue problem, first for the ground state configurations and then for the excited states.

\subsection{Ground state}

\begin{figure*}
	\centering
    \includegraphics[width=18.cm]{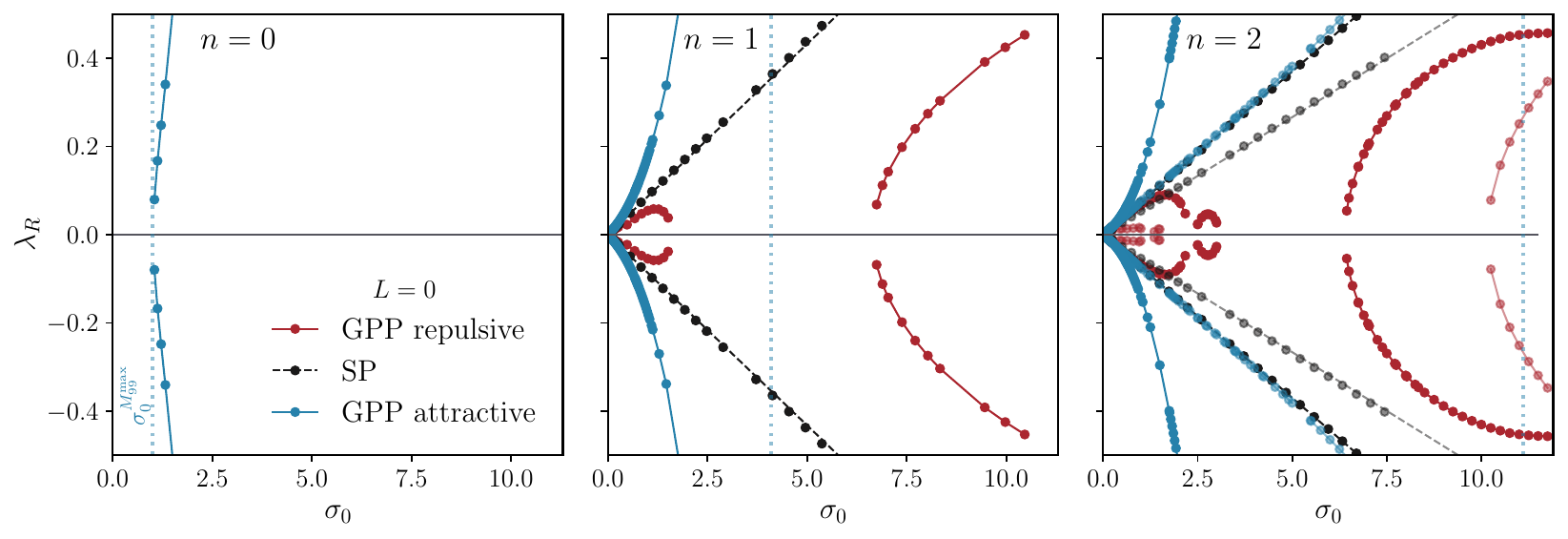}	
    \caption{{\bf Spherically symmetric perturbations:} The real part of the frequency $\lambda$ as a function of the central amplitude $\sigma_0$ for perturbations with $L=0$ in configurations with $n=0$, $1$ and $2$ nodes. The existence of modes with a real positive part of the frequency indicates the existence of an instability. The different color shades  correspond to different unstable modes, and for reference we have included dotted vertical lines in blue to indicate the state of maximum mass $\sigma_0^{M^{\mathrm{max}}_{99}}$ in the attractive case. The circles denote our numerical results, whereas the dashed lines represent the scaled results of the $n$ unstable modes  reported in Ref.~\cite{Nambo:2023yut} for the Schr\"odinger-Poisson system. For the repulsive and the attractive cases we use straight solid lines to connect the numerical results and improve the visualization. Ground state configurations ($n=0$) are stable for any value of the amplitude if the selfinteraction is repulsive, or if there is no selfinteraction, and they are unstable for large enough amplitudes ($\sigma_0\gtrsim\sigma_0^{M^{\mathrm{max}}_{99}}\approx 1.0$) if the selfinteraction is attractive. Excited states ($n\neq 0$) are in general unstable, although certain bands of stability in the amplitude $\sigma_0$ appear if the selfinteraction is repulsive.
 }\label{FigAutovL0}
\end{figure*}

\begin{figure*}
	\centering	
    \includegraphics[width=18.cm]{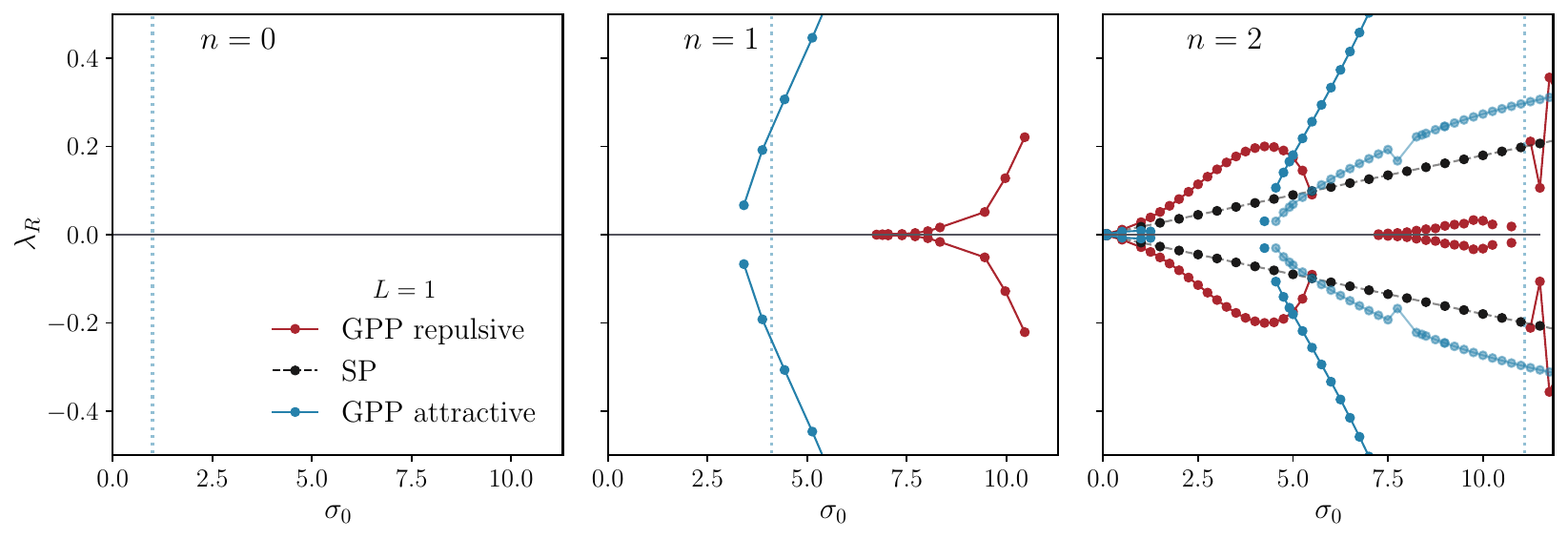}	
    \caption{{\bf Non-spherical perturbations ($L=1$):} Same as in Fig.~\ref{FigAutovL0} but for perturbations with $L=1$. The ground state ($n=0$) is always stable under such perturbations, no matter if the selfinteraction is attractive, repulsive or absent. Excited states ($n\neq 0$) present stability bands for the repulsive and attractive cases.}\label{FigAutovL1}
\end{figure*}

As we have discussed in previous sections, the Gross-Pitaevskii-Poisson system describes three possible scenarios: one characterized by the absence of selfinteractions (where the Gross-Pitaevskii-Poisson equations reduce to the Schr\"odinger-Poisson system), another where the selfinteraction is repulsive, and a third scenario where the selfinteraction is attractive. 

\begin{figure*}
	\centering	
    \includegraphics[width=18.cm]{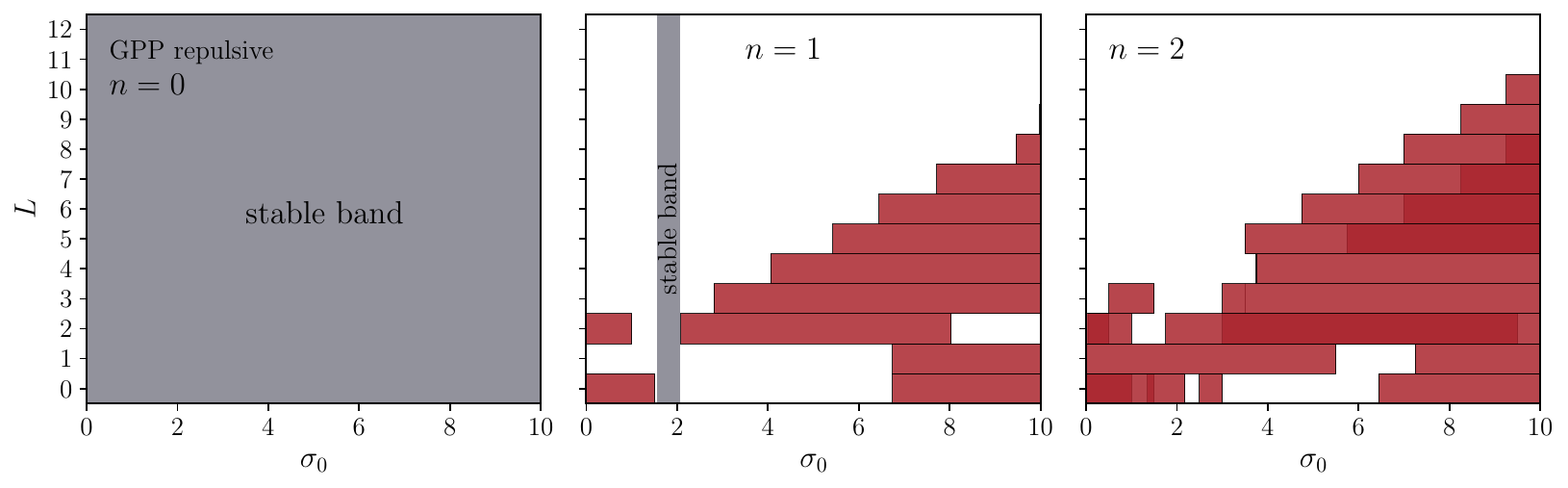}
	\caption{{\bf Stability band structure (repulsive case):} The shaded regions in red represent, for different values of $L$ in the interval from $0$ to $12$, the amplitudes $\sigma_0$ for which the system posses unstable modes. Lighter (darker) colors indicate the presence of one (two) unstable mode(s). Note that the ground state ($n=0$) does not present unstable modes for any value of $\sigma_0$ or $L$, so we can conclude that it is stable. In contrast, the first excited state ($n=1$) presents a common stability band in the interval $1.55\lesssim \sigma_0\lesssim 2.07$, whereas this band is empty for the second excited state ($n=2$).}\label{FigAutovPL}
\end{figure*}

For ground state configurations ($n=0$) we have found that, for the explored range of the parameter space ($0<\sigma_0<10$, $0\le L\le 12$ for the repulsive and $0<\sigma_0<5$, $0\le L\le 6$ for the attractive case), the first two scenarios exhibit a similar behavior under linear perturbations. In particular, the configurations are always stable under spherically symmetric perturbations ($L=0$), as can be appreciated from the left panel of Fig.~\ref{FigAutovL0}. Furthermore, non-spherical perturbations ($L\neq0$) do not either exhibit unstable modes that grow in time, at least for $L\le 12$, as can be seen in the left panel of Fig.~\ref{FigAutovL1} for $L=1$ and the left panel of Fig.~\ref{FigAutovPL} for other values of $L$ in the repulsive case. The results corresponding to the scenario without selfinteractions are consistent with those reported in~\cite{Nambo:2023yut} and coincide with the repulsive and the attractive cases in the limit $\sigma_0\to 0$.

\begin{figure*}
	\centering	
    \includegraphics[width=18.cm]{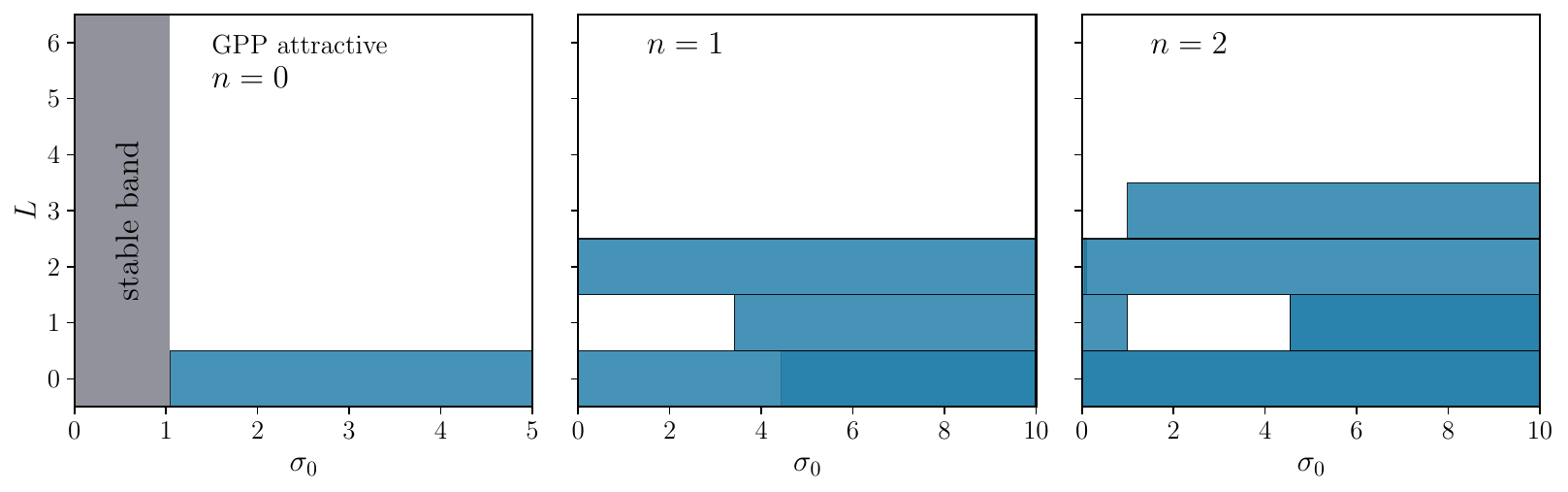}
	\caption{{\bf Stability band structure (attractive case):} Same as in Fig.~\ref{FigAutovPL} but for the attractive case. Note that the common stability band is empty for the first and second excited states and that no unstable modes are found for $L>3$. Note also that for $n=2$ and $L=2$ there are two unstable modes in the interval $0<\sigma_0<0.1$, although they are barely visible.}\label{FigAutovNL}
\end{figure*}

Contrary to the first two scenarios, the case with an attractive selfinteraction presents ground state configurations that are unstable under radial perturbations ($L=0$). More precisely, we found unstable modes with $\lambda_{R}>0$ when 
$\sigma_0\gtrsim 1$; see the left panel of Fig.~\ref{FigAutovL0}. This gives rise to a family of solutions which is stable for $\sigma_0\lesssim 1$ and unstable for $\sigma_0\gtrsim 1$. Interestingly, the division between the stable and unstable states seems to coincide with the maximum mass configuration at $\sigma_0 = \sigma_0^{M_{99}^{\textrm{max}}}\approx 1$.  This is consistent with the turning-point principle; see Refs.~\cite{Zel’dovich63, 1965gtgc.book.....H, Straumann:1984xf}, the Appendix C2 of the published version of~\cite{Alberti:2018lqk}, and in particular~\cite{Gleiser:1988ih} for a discussion in the context of relativistic boson stars. Regarding non-spherical perturbations, no unstable modes appear when $L=1$, as is appreciated from the left panel of Fig.~\ref{FigAutovL1}. Based on a more extensive study within the aforementioned parameter range, we found no unstable modes for $L > 0$, as can be seen from the left panel of Fig.~\ref{FigAutovNL}.

\subsection{Excited states (spherical perturbations)}

The excited states of the Schr\"odinger-Poisson system are known to be unstable under radial perturbations~\cite{2002math.ph...8045H}.
However, in presence of selfinteractions, our analysis (carried out in the range $0<\sigma_0<14$ with $0\le L\le 12$ and $0\le L\le 6$ in the repulsive and attractive cases, respectively) reveals an interesting pattern of stability and instability bands, with a particular relevance for the case where the selfinteraction is repulsive. Our study focusses on the first two excited states ($n=1$ and $n=2$) of the Gross-Pitaevskii-Poisson equations, however, we expect that the behavior discussed below will continue to be valid for higher excited states.

The central and right panels of Fig.~\ref{FigAutovL0} present the results obtained for radial perturbations. As can be observed, there exists unstable modes with $\lambda_R>0$ in the attractive, the repulsive as well as in the non-selfinteracting case. In the first and third cases, we have identified an unstable band that spans all values of the central amplitude $\sigma_0$ that we have explored. For the $n$'th excited state, this band consists of $n$ unstable modes that grow exponentially in time. In the attractive scenario, for one of these modes, $\lambda_R$ grows rapidly as $\sigma_0$ increases, signaling that the solutions have a lifetime that is much shorter than the state with the same central amplitude in the non-selfinteracting theory. Indeed, in absence of selfinteractions, one observes a linear relationship between the real part of the mode(s) frequency $\lambda_R$ and the central amplitude $\sigma_0$. This relationship arises from an inherent scaling freedom in the Schr\"odinger-Poisson system. For further details regarding this symmetry we refer the reader to Refs.~\cite{Moroz:1998dh, Roque:2023sjl}. The dashed lines represent the scaled results of the $n$ unstable modes reported in Ref.~\cite{Nambo:2023yut}, while the dark circles denote our numerical results. As can be seen, both sets of results are in agreement. 

A more interesting pattern appears for the repulsive scenario, in which the unstable radial modes cover only localized intervals of $\sigma_0$, resulting in stability  bands where $\lambda_R=0$;\footnote{Note that the occurrence of such localized stability bands is excluded in the non-selfinteracting case due to the aforementioned rescaling freedom of the Schr\"odinger-Poisson system.} see the central and right panels of Fig.~\ref{FigAutovL0}. It is worth noting that, unlike ground state configurations, the boundary between stable and unstable solutions is not determined by a maximum mass state.  The relation of these results with those of Refs.~\cite{Sanchis-Gual:2021phr, Brito:2023fwr} will be discussed in the conclusion section and Appendix~\ref{app.comparison}.

\subsection{Excited states (non-spherical perturbations)} 

Next, we address the problem of what happens to the stability bands of the previous subsection when non-spherical perturbations are considered. Specifically, the relevant question is whether unstable modes with $L > 0$ appear inside these bands, which would imply that the configurations are unstable with respect to generic linear perturbations.

The central and right panels of Fig.~\ref{FigAutovL1} show  unstable modes with $L=1$ appearing in certain intervals of $\sigma_0$, giving rise to new stability bands. Interestingly, such bands appear in the repulsive as well as in the attractive case, whereas for $L=0$ they only occurred in the former case. Further stability bands arise as $L$ increases. The central and right panels of Figs.~\ref{FigAutovPL} and~\ref{FigAutovNL} show the different stability bands (white regions) for different values of $L$. Notice that for high enough values of $L$, the stable band covers entirely the region of the parameter space we have explored, which is compatible with the result obtained in Sec.~\ref{Sec:IVB}.

Furthermore, it is interesting to compare the colored regions of instability shown in Figs.~\ref{FigAutovPL} and~\ref{FigAutovNL} for small values of $\sigma_0$. In this limit the selfinteraction term of the Gross-Pitaevskii equation~(\ref{eqs.GPP}) is subdominant, which would lead to all scenarios being equivalent (see the overlapping region in Figs.~\ref{FigFondn0} and~\ref{FigFondn1}). As a consequence, one observes from Figs.~\ref{FigAutovPL} and~\ref{FigAutovNL} that in the limit $\sigma_0\to 0$ the instability bands (including the number of unstable modes) transit continuously to the corresponding instability bands belonging to the configurations of the Schr\"odinger-Poisson system. For the particular cases $n=0$ and $n=1$ we refer the reader to Fig.~4 of Ref.~\cite{Nambo:2023yut} for the case $\ell=0$ where the respective regions of instability for the Schr\"odinger-Poisson system are reported.

We conclude this section by emphasizing that only configurations lying in the intersection of the stability bands for all $L$ are stable with respect to generic linear perturbations. For $n=1$ the resulting common stability band turns out to be very narrow in the repulsive case, as shown in Fig.~\ref{FigAutovPL}, whereas it is empty in the attractive case, see Fig.~\ref{FigAutovNL}. For $n=2$ there are no stable configurations, neither in the repulsive nor in the attractive case. Similar results are expected for higher excited states.

\section{Conclusions}

In this paper we have analyzed the impact of a selfinteraction potential on the linear stability of boson stars, with particular interest on the excited states. In absence of selfinteractions, it is known that excited boson stars are unstable, as has been confirmed using semianalytical techniques, both in the nonrelativistic~\cite{2002math.ph...8045H} and the relativistic~\cite{Lee:1988av} regimes, and numerical simulations~\cite{Balakrishna:1997ej, 2002math.ph...8045H}. This has led to the belief that this result is general and holds true in presence of a selfinteraction potential, as has been  observed using numerical evolutions of the spherically symmetric Einstein-Klein-Gordon equations in the $\lambda|\phi|^4$ theory~\cite{Balakrishna:1997ej}. 

However, in a series of recent papers, performing a more systematic study of the numerical evolutions of selfinteracting excited boson star configurations the authors of Refs.~\cite{Sanchis-Gual:2021phr, Brito:2023fwr} have concluded that, for strong enough repulsive quartic selfinteraction, these configurations remain stable under spherical perturbations, at least for times of the order of $10^4m_0^{-1}$. 

Motivated by these recent findings, and using a combination of analytic and numerical methods, we have analyzed the linear stability of the stationary and spherically symmetric solutions of the Gross-Pitaevskii-Poisson system that describes the nonrelativistic limit of the Einstein-Klein-Gordon theory in presence of a quartic or more general potential term. Furthermore, we have been able to extend our analysis to include perturbations that do not necessarily respect the spherical symmetry of the unperturbed system, and in this sense it goes beyond the radial stability study of the excited states performed in~\cite{Sanchis-Gual:2021phr, Brito:2023fwr}.

In particular, if the selfintreaction is repulsive, we have found that ground state configurations are stable under generic linear perturbations, a result that is consistent with the idea that these states represent a global minimum of the energy functional for fixed particle number (cf. the left panel of Fig.~\ref{fig.energNump}). For the excited states, however, there exist a series of stability and instability bands in the central amplitude $\sigma_0$ of the background field that depends on the particular value of the angular momentum number $L$ associated with the perturbation. This series disappears for high enough values of $L$, since in this case there are no unstable modes, as we have been able to prove. The stability under generic linear perturbations is determined by the common band that results from the intersection of the stability bands associated with each different $L$. For the first excited state we have determined the existence of a common stability band in the range $1.55\lesssim \sigma_0\lesssim 2.07$, whereas this band is empty for the second excited state. These results are summarized in Fig.~\ref{FigAutovPL}, and they are independent of the mass of the scalar field and the strength of the selfinteraction (as long as they are different from zero). This is a consequence of the fact that for the Gross-Pitaevskii-Poisson system $m_0$ and $\Lambda$ can be absorbed in the dimensionless variables. We expect a similar behavior for higher excited states.

In contrast, if the selfinteraction is attractive, ground state configurations are only stable if they are located to the right of the maximum mass configuration in the $M_{99}$ vs. $R_{99}$ curve (see the center panel of Fig.~\ref{FigFondn0}). Regarding the excited states, and similar to what occurs in the repulsive case, for each value of the angular momentum number $L$ of the perturbations there exist a series of stability and instability bands in the central amplitude $\sigma_0$.
However, the common stability band is empty, at least for the first two excited states; see Fig.~\ref{FigAutovNL} for details. Again, similar results are expected for higher excited states.

It is reasonable to think that the stability bands we have found persist in the Einstein-Klein-Gordon theory, at least as long as we are not too far from the nonrelativistic limit. Related to this point, it is interesting to compare our results with those reported in Refs.~\cite{Sanchis-Gual:2021phr, Brito:2023fwr}. In~\cite{Sanchis-Gual:2021phr} the authors construct solutions of the Einstein-Klein-Gordon equations for the first excited state ($n=1$) and different values of the selfinteraction parameter $\Lambda$ and frequency $\omega$, whereas in~\cite{Brito:2023fwr}, for the states $n=1, \ldots, 10$, the frequency $\omega$ is fixed and the selfinteraction parameter is varied. In particular, it is found in~\cite{Brito:2023fwr} that for sufficiently large values of $\Lambda$ the excited states are stable under radial perturbations. Additionally, for the case of the first excited states, stability bands can be identified from Fig.~$3$ and Table~$1$ in~\cite{Sanchis-Gual:2021phr}.

Although the existence of excited stable states and the occurrence of stability bands is compatible with the results reported in Refs.~\cite{Sanchis-Gual:2021phr, Brito:2023fwr}, a direct comparison is challenging due to the relativistic effects that we have neglected in our results. Furthermore, the fact that the Gross-Pitaevskii-Poisson system can be rewritten in a form that is independent of the mass and the selfinteraction parameter reduces the problem to one with a single continuous parameter (the field's central amplitude); hence exploring the solution space in the same form as in~\cite{Sanchis-Gual:2021phr, Brito:2023fwr} is unnatural in our setting. For the second excited states, a rough comparison of our results with those of Refs.~\cite{Brito:2023fwr} is presented in Appendix~\ref{app.comparison}, although it is important to stress that, in the nonrelativistic limit, these states are unstable with respect to generic linear perturbations (see the right panel of Fig.~\ref{FigAutovPL} for reference).  

Finally, it is interesting to stress that, contrary to what happens in the relativistic theory, the conclusions that we have presented in this paper are generic and can be extended to other potentials different from the quartic selfinteraction one that have been analyzed in e.g. Refs.~\cite{Balakrishna:1997ej, Sanchis-Gual:2021phr, Brito:2023fwr}. Again, this is due to some of the peculiarities of the nonrelativistic limit of the Einstein-Klein-Gordon theory, as we clarify in Appendix~\ref{app.generic.potential}.
Boson stars resulting from higher-rank fields, both in the absence and presence of self-interactions, have been explored in e.g. Refs.~\cite{Jain:2021pnk,Adshead:2021kvl,Zhang:2021xxa,Jain:2022kwq}.

\begin{acknowledgments}

We are grateful to Nicolas Sanchis-Gual, Marco Brito, Carlos Herdeiro, Eugen Radu and Miguel Zilh\~ao for bringing to our attention the existence of stability/instability bands within the relativistic framework, and to Mudit Jain and Pierre Chavanis for correspondence. This work was partially supported by CONAHCyT Projects No. 376127 ``Sombras, lentes y ondas gravitatorias generadas por objetos compactos astrofísicos'' and No. 286897 ``Materia oscura: Implicaciones de sus propiedades fundamentales en las observaciones astrofísicas y cosmológicas'', and by CONAHCyT-SNII. E.C.N. was supported by a CONAHCyT doctoral scholarship. A.D.T. acknowledges support from DAIP. A.A.R. also acknowledges funding from a postdoctoral fellowship from ``Estancias Posdoctorales por México para la Formación y Consolidación de las y los Investigadores por México''. O.S. was partially supported by a CIC grant to Universidad Michoacana de San Nicolás de Hidalgo. We also acknowledge the use of the computing server COUGHS from the UGDataLab at the Physics Department of Guanajuato University.
\end{acknowledgments}

\appendix

\section{Generic selfinteraction potential}\label{app.generic.potential}

\begin{figure}[t!]
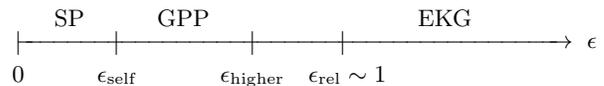

\hspace{-1.15cm}SP  \hspace{0.75cm} GPP \hspace{2.55cm} EKG\\
\vspace{-.75cm}
\[ \xrightarrow{\hspace*{7.2cm}} \; \epsilon \]\\
\vspace{-.65cm}
\hspace{-3.35cm}$|$ \hspace{1cm} $|$ \hspace{1.5cm} $|$\hspace{1cm} $|$\\
\vspace{.06cm}
\hspace{-2.85cm}0 \hspace{0.75cm} $\epsilon_{\textrm{self}}$ \hspace{0.9cm} $\epsilon_{\textrm{higher}}$  \hspace{0.1cm} $\epsilon_{\textrm{rel}}\sim1$\\ 
\caption{{\bf The effect of the higher selfinteraction terms.} The different regimes of the Einstein-Klein-Gordon theory, showing the scales at which the quartic selfinteraction term ($\epsilon_{\textrm{self}}$) and the higher-order terms ($\epsilon_{\textrm{higher}}$) become important. Notice the model independence that the Einstein-Klein-Gordon theory exhibits at scales below $\epsilon_{\textrm{higher}}$. This regime is characterized in terms of two parameters: the mass of the scalar field and the coupling constant $\lambda$. For the quartic selfinteraction potential $\epsilon_{\textrm{higher}}\to\infty$, whereas for a generic potential $\epsilon_{\textrm{higher}}$ is expected to be of order $ \epsilon_{\textrm{rel}}\sim 1$.
}\label{Fig:higher}
\end{figure}

Consider a selfinteraction potential of the general form 
\begin{equation}
\label{eq.general.potential}
V(\phi) = M^4 \sum_{n=2}^{\infty} \frac{v_{2n}}{(2n)!}\left|\frac{\phi}{M}\right|^{2n},
\end{equation}
where $v_{2n}$ are dimensionless constant parameters and $M$ is a characteristic mass scale. Under the assumption that the coefficients $v_{2n}$ do not grow too fast such that the convergence radius is different from zero, this is the most general potential which respects the internal $U(1)$ symmetry and is analytic in $|\phi|^2$ in a vicinity of $\phi=0$. Furthermore, we have assumed that the constant term $n=0$ is zero and have excluded the $n=1$ term in Eq.~(\ref{eq.general.potential}) since its contribution is already included in the mass of the scalar field $m_0$. Note that the first term of this series corresponds to the quartic $\lambda|\phi|^4$ selfinteraction potential of Eq.~(\ref{eq.action2}), with $\lambda=v_4/4!$; however, now we have an infinite number of additional selfinteraction terms that also contribute to the potential energy of the field. 

After introducing this potential into the action~(\ref{eq.action}) and working out the nonrelativistic approximation as we described in Sec.~\ref{sec:nonrelativistic}, the selfinteraction term in Eq.~(\ref{eq.action.nonrel}) is modified to
\begin{equation}
-\frac{\lambda}{4m_0^2}|\psi|^4\left[1+\frac{v_6}{240v_4}\frac{|\psi|^2}{m_0M^2}+\ldots\right].
\end{equation}
In the nonrelativistic limit $\psi\sim \sqrt{M_{\textrm{Pl}}^2m_0}\epsilon$, and the second term in the square bracket starts to contribute to the equations of motion when $\epsilon\sim\epsilon_{\textrm{higher}}= \sqrt{240v_4/v_6}(M/M_{\textrm{Pl}})$, which should be compared with $\epsilon_{\textrm{self}}= m_0^2/(\lambda M_{\textrm{Pl}}^2)$, that signals the onset of the quartic selfinteraction and is naturally smaller. For instance, if $M\sim M_{\textrm{Pl}}$ and all the coefficients $v_{2n}$ are of the same order then the higher orders of the potential show up at the same scale than the relativistic effects, when $\epsilon\sim\epsilon_{\textrm{rel}}= 1$. This leads to the conclusion that the nonrelativistic limit that we have introduced in Eq.~(\ref{eq.action.nonrel}) is generic and valid beyond the quartic selfinteraction theory $\lambda|\phi|^4$. These observations are illustrated pictorially in Fig.~\ref{Fig:higher}.

\section{Comparison with the work in Refs.\texorpdfstring{~\cite{Brito:2023fwr, Sanchis-Gual:2021phr}}{}}\label{app.comparison}

The Einstein-Klein-Gordon action in Eq.~(\ref{eq.action}) of our paper coincides with that of Eq.~(1) in Refs.~\cite{Sanchis-Gual:2021phr, Brito:2023fwr} provided we perform the following identifications: $M_{\textrm{Pl}}=1$, $m_0=\mu$ and $\phi=\frac{1}{\sqrt{2}}\Phi$. However, there is a factor $2$ of difference in our definition of the selfinteraction parameter~(\ref{eq.Lambda}), i.e. $\Lambda_{\textrm{ours}}=2\Lambda_{\textrm{their}}$. Taking this into account, it is possible to conclude that, in the nonrelativistic limit (cf. Eqs.~(\ref{eq.code.numbers1}) and~(\ref{eq.harmonic})),
\begin{equation}
\bar{\sigma}_0 = \left(\frac{\pi\Lambda_{\textrm{ours}}^2}{2M_{\textrm{Pl}}^2m_0}\right)^{1/2} \sigma_0 = \sqrt{2\pi}\Lambda_{\textrm{their}}\frac{|\Phi_0|}{M_{\textrm{Pl}}},
\end{equation}
where in this appendix we have reintroduced the overbar to indicate dimensionless quantities. Given that the authors of Refs.~\cite{Sanchis-Gual:2021phr, Brito:2023fwr} work in Planck units, $M_{\textrm{Pl}}=1$, we can relate field amplitudes in our paper to those of Refs.~\cite{Sanchis-Gual:2021phr, Brito:2023fwr} through $\bar{\sigma}_0 =\sqrt{2\pi}\Lambda_{\textrm{their}}|\Phi_0|$.

In Ref.~\cite{Brito:2023fwr} it is reported that, for second excited states ($n=2$) with frequency $\omega=0.92$ and $\mu=1$,  the minimum value of the selfinteraction parameter for which a configuration is stable under radial perturbations is $\Lambda_{\textrm{their}}=150$, corresponding to a central amplitude of approximately $\Phi_0 \approx 1.3 \times 10^{-2}$ (see Fig.~7 in~\cite{Brito:2023fwr}), which translates to $\bar{\sigma}_0 \approx 4.9$ in our paper. We can compare this result with the stability bands at $L=0$ that we have identified in the right panel of Fig.~\ref{FigAutovPL}, i.e. $2.2<\bar{\sigma}_0<2.5$ and $3<\bar{\sigma}_0<6.4$. Even if the correspondence is only at the order of magnitude level, it is important to stress that the configurations with $\omega=0.92$ are relativistic, so we do not expect a perfect match.

Furthermore, evidence of stability/instability bands can be also inferred from the results presented in Ref~\cite{Sanchis-Gual:2021phr}. In Fig.~3 of this reference, the authors explore the solution space corresponding to the first excited states ($n = 1$) by varying the selfinteraction parameter $\Lambda$ and the frequency $\omega$. For instance, choosing $\Lambda = 75$, a stability band emerges as we decrease the frequency from $\omega = 0.92$ to $\omega=0.90$, leading to excited states that are stable under radial perturbations. However, further decrease in $\omega$ leads again to unstable solutions. Table~$1$ of this reference confirms the same pattern.

\bibliography{ref.bib} 

\end{document}